\documentclass[epj]{svjour}
\usepackage{amscd}
\usepackage{amssymb}
\usepackage{amsmath}
\usepackage[T1]{fontenc}
\usepackage[latin1]{inputenc}
\usepackage{graphicx}
\usepackage{graphics}
\usepackage{color}
\usepackage{latexsym}
\usepackage{amsfonts}
\def\be{\begin{equation}}
\def\ee{\end{equation}}
\def\ba{\begin{eqnarray}}
\def\ea{\end{eqnarray}}
\def\>{\rangle}
\def\<{\langle}
\def\n{\nonumber}

\def\sc{\scriptsize}

\begin{document}
\title{Jarzynski equality and the second law of thermodynamics beyond the weak-coupling limit: The quantum Brownian oscillator}
\author{Ilki Kim\thanks{\emph{e-mail}: hannibal.ikim@gmail.com}}
\institute{Center for Energy Research and Technology, North Carolina
A$\&$T State University, Greensboro, North Carolina 27411, USA}
\date{\today}
%
%%%%%%%%%%%%%%%%%%%%%%%%%%%%%%%%%%%%%%%%%%%%%%%%%%%%%%%%%%%%%%%%%%%%%%%%%%%%%
\abstract{We consider a time-dependent quantum linear oscillator
coupled to a bath at an arbitrary strength. We then introduce a
generalized Jarzynski equality (GJE) which includes the terms
reflecting the system-bath coupling. This enables us to study
systematically the coupling effect on the linear oscillator in a
non-equilibrium process. This is also associated with the second law
of thermodynamics beyond the weak-coupling limit. We next take into
consideration the GJE in the classical limit. By this generalization
we show that the Jarzynski equality in its original form can be
associated with the second law, in both quantal and classical
domains, only in the vanishingly small coupling regime.
 % end of PACS codes
}
\authorrunning{I. Kim}
\titlerunning{Jarzynski equality and the second law of thermodynamics $\cdots$}
\maketitle
%
%%%%%%%%%%%%%%%%%%%%%%%%%%%%%%%%%%%%%%%%%%%%%%%%%%%%%%%%%%%%%%%%%%%%%%%%%%%%%
\section{Introduction}\label{sec:introduction}
%%%%%%%%%%%%%%%%%%%%%%%%%%%%%%%%%%%%%%%%%%%%%%%%%%%%%%%%%%%%%%%%%%%%%%%%%%%%%
Since it was introduced, the Jarzynski equality (JE) has been
attracting a great deal of interest due to its remarkable attribute.
It explicitly states that if a given system, initially prepared in
thermal equilibrium, is driven far from this equilibrium by an
external perturbation, then this non-equilibrium process satisfies
\cite{JAR97}
\begin{equation}\label{eq:jarzynski1}
    \left\<e^{-\beta\, W(t_{\text{\tiny f}})}\right\>\, =\, \int dW\cdot P(W)\cdot e^{-\beta\, W(t_{\text{\tiny f}})}\, =\, e^{-\beta\, \Delta F}\,,
\end{equation}
where the symbol $W(t_{\text{\sc f}})$ denotes the microscopic work
performed on the system in a single run for a variation of the
external parameter $\{\lambda(t)\,|\,0 \leq t \leq t_{\text{\sc
f}}\}$ according to the pre-determined protocol, and the symbol
$P(W)$ is the probability distribution of the work value $W$; the
symbol $\beta = 1/(k_{\mbox{\tiny B}} T)$ denotes the inverse
temperature of the environment, and $\Delta F$ is the Helmholtz free
energy change of the system between the initial and final states,
equivalent to the average reversible work $\<W\>_{\text{\sc rev}}$
in the corresponding isothermal process \cite{CAL85}. As such, the
average $\<\cdots\>$ is evaluated over a large number of runs. Then,
applying the Jensen inequality to the JE (\ref{eq:jarzynski1}), we
can easily obtain $\<W\> \geq \Delta F$ as an expression of the
second law of the thermodynamics \cite{JAR08,BOK10}. Further, as a
generalization of JE, Crooks introduced the fluctuation theorem
given by \cite{CRO98,CRO99}
\begin{equation}\label{eq:haenggi-fluctuation_1}
    \frac{P_{\text{\sc f}}(+W)}{P_{\text{\sc r}}(-W)}\; =\; \exp\{\beta\,(W - \Delta F)\}\,,
\end{equation}
where the symbols $P_{\text{\sc f}}(W)$ and $P_{\text{\sc r}}(W)$
are the probability densities of performing the work $W$ when its
protocol runs in the forward and reverse directions, respectively.

The JE has been verified in experiments with small-scale systems,
where the fluctuations of work values are sufficiently observable
\cite{JAR02}, e.g., in determination of the free energy change
between the unfolded and folded conformations of a single RNA
molecule \cite{LIP02,BUS05}. However, its strict validity is still
in dispute, especially in association with the second law beyond the
weak coupling between system and bath (see, e.g., an instructive
critique in \cite{COH04} as well as Jarzynski's reply in
\cite{JAR04}).

The JE has also been discussed in the quantum domain, where no
notion of trajectory in the phase space is available and so instead
the spectrum information has to be used for determination of the
work probability distribution $P(W)$. Most of those attempts were
made within isolated or weakly-coupled systems
\cite{TAS00,MUK03,ROE04,ESP06,TAL07,MAH07,ENG07,CRO08,AND08,DEF08,TAK09,MAH10,MAH11,NGO12}.
However, the finite coupling strength between system and bath in
small-scale devices gives rise to some quantum subtleties which can
no longer be neglected for studying the thermodynamic properties of
such devices \cite{SHE02,MAH04,CAP05}.

In \cite{CAM09}, the fluctuation theorem
(\ref{eq:haenggi-fluctuation_1}), immediately reproducing the JE,
was discussed for arbitrary open quantum systems with no restriction
on the coupling strength. Its key idea was such that if the total
system (i.e., open quantum system plus bath) is initially in a
thermal canonical state, but otherwise isolated, and an external
perturbation $\lambda(t)$ acts solely on the open system, then the
work performed on the total system may be interpreted as the work
performed on the open system alone. Then the final result was
explicitly obtained [cf.
(\ref{eq:haenggi-crooks-fluctuation_1-1})-(\ref{eq:haenggi-jarzynski_1-1})]
by mimicking the classical case such that the work on the
system-of-interest $H_s(x,\lambda)$ beyond the weak-coupling limit
is directly related to the free energy change of a potential of mean
force \cite{JAR04}
\begin{eqnarray}\label{eq:jarzynski_form1}
    {\mathcal H}_s^*(x,\lambda) &=& H_s(x,\lambda)\, -\, \beta^{-1} \times\\
    && \ln\left(\frac{\int dy\, \exp\left[-\beta\, \{H_{b}(y)\, +\, h_{\text{\tiny int}}(x,y)\}\right]}{\int dy\, \exp\left\{-\beta\, H_{b}(y)\right\}}\right)\,,\n
\end{eqnarray}
where the symbols $H_b(y)$ and $h_{\text{\tiny int}}(x,y)$ denote
the bath and interaction Hamiltonians, respectively; note here that
on the right-hand side the $\lambda$-dependency exists only in
$H_s(x,\lambda)$ [cf. (\ref{eq:zassenhaus2})]. This quantum result
was subsequently applied to a solvable model (e.g., \cite{TAL09}).

However, as will be discussed below, it is not clear if this quantum
fluctuation theorem is associated directly with the
system-of-interest $\hat{H}_s(t)$ alone, beyond the weak-coupling
limit, being the quantum description of $H_s(x,\lambda)$ in
(\ref{eq:jarzynski_form1}); in fact, an attempt to relate the
quantum description of ${\mathcal H}_s^*(x,\lambda)$ to the second
law of thermodynamics within the coupled system $\hat{H}_s(t)$ would
even lead to a violation of this law. In this paper we intend to
resolve this issue, within a time-dependent quantum Brownian
oscillator as a mathematically manageable scheme, by introducing
explicitly a generalized Jarzynski equality (GJE) directly
associated with the open system $\hat{H}_s(t)$ with no restriction
on the coupling strength, and then discussing the resultant second
law with no violation. The general layout of this paper is the
following. In Sect. \ref{sec:basics} we briefly review the results
known from the references and useful for our discussions. In Sect.
\ref{sec:2nd_law_drude} we discuss the second law beyond the
weak-coupling regime, which shows that the JE in its known form can
be associated with this law only in the vanishingly small coupling
regime. In Sect. \ref{sec:Jarzynski} we introduce the GJE consistent
with the second law, valid at an arbitrary coupling strength.
Finally we give the concluding remarks of this paper in Sect.
\ref{sec:conclusions}.

%%%%%%%%%%%%%%%%%%%%%%%%%%%%%%%%%%%%%%%%%%%%%%%%%%%%%%%%%%%%%%%%%%%%%%%%%%%%%
\section{Quantum Brownian oscillator in a non-equilibrium thermal process}\label{sec:basics}
%%%%%%%%%%%%%%%%%%%%%%%%%%%%%%%%%%%%%%%%%%%%%%%%%%%%%%%%%%%%%%%%%%%%%%%%%%%%%
%
The quantum Brownian oscillator in consideration is given by the
Caldeira-Leggett model Hamiltonian \cite{ING98,WEI08}
\begin{equation}\label{eq:total_hamiltonian1}
    \hat{H}(t)\; =\; \hat{H}_s(t)\, +\, \hat{H}_b\, +\, \hat{H}_{sb}\,,
\end{equation}
where a system linear oscillator, a bath, and a system-bath
interaction are given by
\begin{align}
    \hat{H}_s(t)\; &=\; \frac{\hat{p}^2}{2 M}\,+\,\frac{M\,y^2(t)}{2}\,\hat{q}^2\tag{\ref{eq:total_hamiltonian1}a}\label{eq:total_hamiltonian2-1}\\
    \hat{H}_b\; &=\; \sum_{j=1}^N \left(\frac{\hat{p}_j^2}{2 m_j}\,+\,\frac{m_j\,\omega_j^2}{2}\,\hat{x}_j^2\right)\tag{\ref{eq:total_hamiltonian1}b}\label{eq:total_hamiltonian2-2}\\
    \hat{H}_{sb}\; &=\; -\hat{q} \sum_{j=1}^N c_j\,\hat{x}_j\,+\,\hat{q}^2
    \sum_{j=1}^N \frac{c_j^2}{2 m_j\,\bar{\omega}_j^2}\,,\tag{\ref{eq:total_hamiltonian1}c}\label{eq:total_hamiltonian2-3}
\end{align}
respectively. Here the angular frequency $y(t) > 0$ varies in the
time interval $[0, t_{\text{\sc f}}]$ according to an arbitrary but
pre-determined protocol (for the sake of convenience, let $y_0 =
y(0)$ in what follows), and the constants $c_j$ denote the coupling
strengths. The total system $\hat{H}(0)$ initially prepared is in a
canonical thermal equilibrium state $\hat{\rho}_{\beta} = e^{-\beta
\hat{H}(0)}/Z_{\beta}(y_0)$ with the initial partition function
$Z_{\beta}(y_0)$. Then the initial internal energy of the coupled
oscillator is given by
$\mbox{Tr}\{\hat{H}_s(0)\,\hat{\rho}_{\beta}\} =
\mbox{Tr}_s\{\hat{H}_s(0)\,\hat{R}_s(0)\}$, where the initial state
of the oscillator $\hat{R}_s(0) = \mbox{Tr}_b(\hat{\rho}_{\beta})$.
Here the symbol $\mbox{Tr}_b$ denotes the partial trace for the bath
degrees of freedom only; it is explicitly given by
\cite{GRA88,WEI08}
\begin{eqnarray}\label{eq:density_operator1}
    \<q|\hat{R}_s(0)|q'\> &=& \frac{1}{\sqrt{2\pi \<\hat{q}^2\>_{\beta}}}\, \times\\
    && \exp\left\{-\frac{(q + q')^2}{8\,\<\hat{q}^2\>_{\beta}} -
    \frac{\<\hat{p}^2\>_{\beta}\cdot(q - q')^2}{2 \hbar^2}\right\}\,,\n
\end{eqnarray}
expressed in terms of the equilibrium fluctuations,
$\<\hat{q}^2\>_{\beta} = \mbox{Tr} (\hat{q}^2 \hat{\rho}_{\beta})$
and $\<\hat{p}^2\>_{\beta} = \mbox{Tr} (\hat{p}^2
\hat{\rho}_{\beta}) = M^2 \<\dot{\hat{q}}^2\>_{\beta}$. Beyond the
weak-coupling limit, this reduced density matrix is not any longer
in form of a canonical thermal state $\propto e^{-\beta
\hat{H}_s(0)}$, immediately leading to the fact that the coupled
oscillator $\hat{H}_s(0)$ is not with its well-defined local
equilibrium temperature \cite{KIM10}.

For the below discussions, we will need the fluctuation-dissipation
theorem in the initial (equilibrium) state \cite{FOR88}
\begin{eqnarray}\label{eq:dissipation_fluctuation_theorem}
    \hspace*{-.2cm}&&\frac{1}{2}\, \left\<\hat{q}(t_1)\,\hat{q}(t_2)\, +\,
    \hat{q}(t_2)\,\hat{q}(t_1)\right\>_{\beta}\; =\; \frac{\hbar}{\pi}\, \times\\
    \hspace*{-.2cm}&&\int_0^{\infty} d\omega\,
    \coth\left(\frac{\beta \hbar \omega}{2}\right)\; \cos\{\omega(t_2 -
    t_1)\}\; \mbox{Im}\{\tilde{\chi}(\omega + i\,0^+)\}\,.\n
\end{eqnarray}
Here the dynamic susceptibility is given by
\begin{equation}\label{eq:chi_tilde1}
    \tilde{\chi}(\omega)\; =\; \frac{1}{M}\,\frac{1}{y_0^2 -\omega^2 -
    i \omega\,\tilde{\gamma}(\omega)}\,,
\end{equation}
where the symbol $\tilde{\gamma}(\omega)$ denotes the
Fourier-Laplace transformed damping kernel. This can be rewritten as
\cite{LEV88}
\begin{align}\tag{\ref{eq:chi_tilde1}a}\label{eq:susceptibility2}
    \tilde{\chi}(\omega)\; =\; -\frac{1}{M}\, \frac{\displaystyle
    \prod_{j=1}^N\, (\omega^2 - \omega_j^2)}{\displaystyle \prod_{k=0}^N\, (\omega^2 -
    \bar{\omega}_k^2)}
\end{align}
in terms of the normal-mode frequencies $\{\bar{\omega}_k\}$ of the
total system $\hat{H}(0)$. It is straightforward to verify that
$\mbox{Im}\{\tilde{\chi}(\omega + i0^+)\} \to \pi/(2 M
y_0)\cdot\delta(\omega - y_0)$ for an uncoupled (or isolated)
oscillator (i.e., all system-bath coupling strengths $c_j \equiv
0$).

From (\ref{eq:dissipation_fluctuation_theorem}), it follows that
\cite{FOR85}
\begin{subequations}
\begin{eqnarray}
    \hspace*{-.7cm}\<\hat{q}^2\>_{\beta} &=& \frac{\hbar}{\pi}\,\int_0^{\infty} d\omega\,
    \coth\left(\frac{\beta \hbar \omega}{2}\right)\, \mbox{Im}\{\tilde{\chi}(\omega + i\,0^+)\}\label{eq:x_correlation1}\\
    \hspace*{-.7cm}\<\dot{\hat{q}}^2\>_{\beta} &=& \frac{\hbar}{\pi}\,\int_0^{\infty} d\omega\,
    \omega^2\,\coth\left(\frac{\beta \hbar \omega}{2}\right)\, \mbox{Im}\{\tilde{\chi}(\omega + i\,0^+)\}\,,\label{eq:x_dot_correlation1}
\end{eqnarray}
\end{subequations}
as well-known, both of which give the initial internal energy of the
coupled oscillator
\begin{eqnarray}\label{eq:energy1}
    &&U_s(0)\; :=\; \<\hat{H}_s(0)\>_{\beta}\; =\; \frac{M \hbar}{2 \pi}\, \times\\
    &&\int_0^{\infty} d\omega\, (y_0^2 + \omega^2)\,
    \coth\left(\frac{\beta \hbar \omega}{2}\right)\cdot\mbox{Im}\{\tilde{\chi}(\omega + i\,0^+)\}\,.\n
\end{eqnarray}
In the absence of the system-bath coupling, this reduces to the
well-known expression of internal energy \cite{CAL85}
\begin{align}\tag{\ref{eq:energy1}a}\label{eq:internal_energy1}
    e(\beta,y_0)\; =\; \hbar y_0 \left(\frac{1}{2}\, +\,
    \<\hat{n}\>_{\beta}\right)\; =\; \frac{\hbar y_0}{2}\,\coth\left(\frac{\beta \hbar
    y_0}{2}\right)\,,
\end{align}
where the average quantum number $\<\hat{n}\>_{\beta} = 1/(e^{\beta
\hbar y_0} - 1)$. With $\beta \hbar \to 0$, its classical value also
appears as $e_{\text{\sc cl}}(\beta) = 1/\beta$.

In comparison, there is a widely used alternative approach to the
thermodynamic energy of the coupled oscillator in equilibrium, given
by ${\mathcal U}_s^*(0) := -(\partial/\partial \beta)\,\ln {\mathcal
Z}_{\beta}^*(y_0)$ \cite{FOR05,HAE05,HAE06,FOR06,KIM06,KIM07}; the
symbol ${\mathcal Z}_{\beta}^*(y_0)$ denotes the reduced partition
function associated with the Hamiltonian of mean force
\cite{CAM09,JAR04}
\begin{equation}\label{eq:hamiltonian_mean_force1}
    \left.\hat{{\mathcal H}}_s^*(t)\right|_{t=0}\; :=\; -\frac{1}{\beta}\,
    \left.\ln\left[\frac{\mbox{Tr}_b\{e^{-\beta \hat{H}(t)}\}}{\mbox{Tr}_b(e^{-\beta
    \hat{H}_b})}\right]\right|_{t=0}\,,
\end{equation}
and thus ${\mathcal Z}_{\beta}^*(y_0) = \mbox{Tr}_s\{e^{-\beta
\hat{{\mathcal H}}_s(0)}\} = Z_{\beta}(y_0)/(Z_b)_{\beta}$ with the
partition function $(Z_b)_{\beta}$ associated with the isolated bath
$\hat{H}_b$. With the vanishing coupling strengths ($c_j = 0$), this
partition function precisely reduces, as required, to its standard
form of $z_{\beta}(y_0) = \{\mbox{csch}(\beta \hbar y_0/2)\}/2$ for
an isolated oscillator. Then it can be shown that \cite{HAE06}
\begin{subequations}
\begin{equation}\label{eq:compare-partition-function-energy1}
    {\mathcal U}_s^*(0)\; =\; \frac{1}{\beta}\,\left\{1 + \sum_{n=1}^{\infty}
    \frac{2\,y_0^2 +
    \nu_n\,\hat{\gamma}(\nu_n) - \nu_n^2\,\hat{\gamma}'(\nu_n)}{\nu_n^2 +
    \nu_n\,\hat{\gamma}(\nu_n) + y_0^2}\right\}
\end{equation}
with $\nu_n = 2\pi n/\beta \hbar$ and $\hat{\gamma}(z) =
\tilde{\gamma}(iz)$, whereas
\begin{equation}\label{eq:compare-internal-energy1}
    U_s(0)\; =\; \frac{1}{\beta}\,\left\{1 + \sum_{n=1}^{\infty} \frac{2\,y_0^2 +
    \nu_n\,\hat{\gamma}(\nu_n)}{\nu_n^2 + \nu_n\,\hat{\gamma}(\nu_n) + y_0^2}\right\}\,.
\end{equation}
\end{subequations}
It is seen that ${\mathcal U}_s^*(0) \ne U_s(0)$ unless the damping
model is Ohmic. In fact, all thermodynamic quantities resulting from
the partition function ${\mathcal Z}_{\beta}^*(y_0)$ cannot exactly
describe the well-defined thermodynamics of the reduced system
$\hat{H}_s(0)$ in its state (\ref{eq:density_operator1}) beyond the
weak-coupling limit \cite{KIM10}; see also \cite{ING12} for
interesting discussions of the different behaviors between
$\hat{{\mathcal H}}_s^*(0)$ and $\hat{H}_s(0)$ in terms of the
specific heat, within a free damped quantum particle given by
$\hat{H}(t)$ with $y(t) \equiv 0$ in (\ref{eq:total_hamiltonian1}).

It is also instructive to consider a quasi-static (or reversible)
process briefly, which the system of interest undergoes change
infinitely slowly throughout. Then the coupled oscillator remains in
equilibrium exactly in form of Eq. (\ref{eq:density_operator1}) in
every single step such that
\begin{equation}\label{eq:density_operator10}
    \<q|\hat{R}_{\text{\sc eq}}\{y(t)\}|q'\>\; =\; \left.\<q|\hat{R}_s(0)|q'\>\right|_{y_0 \to y(t)}\,.
\end{equation}
Here the time $t$ is understood merely as a parameter specifying the
frequency value $y(t)$. Accordingly, the thermodynamic energy ${\mathcal
U}_s^*(t)$ turns out to be in form of
(\ref{eq:compare-partition-function-energy1}) with $y(t) \leftarrow
y_0$. For later discussions, we rewrite it as \cite{FOR85,KIM07}
\begin{equation}\label{eq:energy_finite_temp1-1}
    {\mathcal U}_s^*(t)\, =\, \frac{1}{\pi} \int_0^{\infty} d \omega\,
    e(\beta,\omega)\cdot\mbox{Im}\left\{\frac{d}{d\omega}\,\ln \tilde{\chi}_t(\omega + i
    0^+)\right\}\,,
\end{equation}
where the second factor of the integrand
\begin{align}
    &\; \mbox{Im}\left\{\frac{d}{d\omega}\,\ln \tilde{\chi}_t(\omega + i
    0^+)\right\}\n\\
    =&\; \pi \left\{\sum_{k=0}^N \delta(\omega - \bar{\omega}_{k,t})\,
    -\, \sum_{j=1}^N \delta(\omega - \omega_j)\right\}\,.\tag{\ref{eq:energy_finite_temp1-1}a}\label{eq:energy_finite_temp1-1-1}
\end{align}
Here the susceptibility $\tilde{\chi}_t(\omega)$ is given by
(\ref{eq:chi_tilde1}) with $y(t) \leftarrow y_0$, as well as
$\{\bar{\omega}_{k,t}\} = \{\bar{\omega}_{k}\}_{y_0 \to y(t)}$. In
the absence of the system-bath coupling, Eq.
(\ref{eq:energy_finite_temp1-1-1}) reduces to $\pi\,\delta\{\omega -
y(t)\}$, as required. Likewise, the free energy defined as
${\mathcal F}_s^*(t) := -(1/\beta)\,\ln {\mathcal
Z}_{\beta}^*\{y(t)\}$, being the total system free energy minus the
bare bath free energy, can also be expressed as \cite{FOR85}
\begin{equation}\label{eq:free_energy_finite_temp1}
    {\mathcal F}_s^*(t)\, =\, \frac{1}{\pi} \int_0^{\infty} d \omega\,
    f(\beta,\omega)\cdot\mbox{Im}\left\{\frac{d}{d\omega}\,\ln \tilde{\chi}_t(\omega + i
    0^+)\right\}\,,
\end{equation}
where the free energy of an isolated oscillator
\begin{align}\tag{\ref{eq:free_energy_finite_temp1}a}\label{eq:free_energy_finite_temp2}
    f(\beta,\omega)\; =\; \frac{\hbar \omega}{2}\, +\, \frac{1}{\beta}\, \ln\left(1 -
    e^{-\beta \hbar \omega}\right)\,,
\end{align}
with $f_{\text{\sc cl}}(\beta,\omega) = \{\ln(\beta \hbar
\omega)\}/\beta$ in the classical limit.

Further, to see explicitly the different behaviors of
(\ref{eq:hamiltonian_mean_force1}) from its classical value
(\ref{eq:jarzynski_form1}), we now apply to the exponentiated
negative total Hamiltonian $\hat{H}(t)$ in
(\ref{eq:hamiltonian_mean_force1}) the Zassenhaus formula with
$\hat{X} := \hat{H}_s(t)$ and $\hat{Y} := \hat{H}_b + \hat{H}_{sb}$
\cite{WIL67},
\begin{equation}\label{eq:zassenhaus1}
    e^{s\,(\hat{X} + \hat{Y})}\; =\; e^{s\,\hat{X}}\cdot
    e^{s\,\hat{Y}}\cdot e^{s^2\,\hat{C}_2}\cdot e^{s^3\,\hat{C}_3}\, \cdots
\end{equation}
where $\hat{C}_2 = -(1/2)\,[\hat{X}, \hat{Y}]$, and $\hat{C}_3 =
-(2/3)\,[\hat{Y}, \hat{C}_2] -(1/3)\,[\hat{X}, \hat{C}_2]$; this
enables Eq. (\ref{eq:hamiltonian_mean_force1}) to be rewritten as
\begin{equation}\label{eq:zassenhaus2}
    \hat{\mathcal H}_s^*(t)\; =\; \hat{H}_s(t)\, -\,
    \frac{1}{\beta}\,\ln\left\{\frac{e^{-\beta\,\hat{Y}}\, e^{\beta^2\,\hat{C}_2}\, e^{-\beta^3\,\hat{C}_3(t)}\,
    \cdots}{\mbox{Tr}_b(e^{-\beta\,\hat{H}_b})}\right\}\,.
\end{equation}
Here the operator $\hat{C}_3 = \hat{C}_3(t)$, and so the second term
on the right-hand side is time-dependent, which is not the case in
its classical value (\ref{eq:jarzynski_form1}), to be noted for our
discussions below.

An extension of the Crooks fluctuation theorem
(\ref{eq:haenggi-fluctuation_1}) to arbitrary open quantum systems
was introduced in \cite{CAM09}, which is valid regardless of the
system-bath coupling strength; this reads as
\begin{equation}\label{eq:haenggi-crooks-fluctuation_1-1}
    \frac{P_{\text{\sc f}}(+W)}{P_{\text{\sc r}}(-W)}\; =\; e^{\beta\,(W - \Delta{\mathcal F}_s^*)}\,,
\end{equation}
where $\Delta{\mathcal F}_s^*(t_{\text{\sc f}}) = {\mathcal
F}_s^*(t_{\text{\sc f}}) - {\mathcal F}_s^*(0)$, explicitly given in
(\ref{eq:free_energy_finite_temp1}) for a coupled oscillator. This
fluctuation theorem leads to the Jarzynski equality
\begin{align}\tag{\ref{eq:haenggi-crooks-fluctuation_1-1}a}\label{eq:haenggi-jarzynski_1-1}
    \<e^{-\beta\,W}\>_{\scriptscriptstyle{P}_{\text{\tiny f}}(W)}\; =\; e^{-\beta\,\Delta{\mathcal F}_s^*}\,,
\end{align}
and it follows that the average work performed by the external
perturbation satisfies $\<W\> \geq \Delta{\mathcal F}_s^*$. However,
it is normally non-trivial to determine the work distribution
$P_{\text{\sc f}}(W)$ explicitly, which can be obtained only from
the spectrum information of the (isolated) total system
$\hat{H}(t)$. Further, from the time-dependent behavior of
(\ref{eq:zassenhaus2}) it is not apparent whether the free energy
change $\Delta{\mathcal F}_s^*$ is precisely tantamount to the
minimum work on the system-of-interest $\hat{H}_s(t)$ only,
especially beyond the weak-coupling limit.

For later purposes, we also point out that an explicit expression of
the work distribution is known for an isolated oscillator as
\cite{TAL07,DEF08}
\begin{eqnarray}
    P\{W(t_{\text{\sc f}},y_0)\}\, =\, \sum_{m,n}&&\delta\left(W - [E_m\{y(t_{\text{\sc f}})\} -
    E_n(y_0)]\right) \times\n\\
    &&P_{m,n}\{y(t_{\text{\sc f}})\}\cdot P_n(y_0)\label{eq:probability_1-1}
\end{eqnarray}
with the microscopic work $W(t_{\text{\sc f}},y_0)$ in a single
event of the variation $\{y_0 \to y(t_{\text{\sc f}})\}$. Here
$E_n\{y(t)\} = \hbar y(t)\,(n + 1/2)$ is the instantaneous energy
eigenvalue, and the symbol $P_n(y_0) =
e^{-\beta\,E_n(y_0)}/z_{\beta}(y_0)$ denotes the probability of the
occupation number in the initial preparation at temperature $T$, as
well as $P_{m,n}\{y(t_{\text{\sc f}})\}\; =\; |\<\psi_m(t_{\text{\sc
f}})|\hat{U}(t_{\text{\sc f}})|\psi_n(0)\>|^2$ the transition
probability between initial and final states $n$ and $m$, expressed
in terms of the instantaneous eigenstates $|\psi_n(t)\>$ and the
time-evolution $\hat{U}(t_{\text{\sc f}})$ of the time-dependent
harmonic oscillator. Then the JE is explicitly given by
$\<e^{-\beta\,W}\>_{\scriptscriptstyle{P}} = e^{-\beta\,\Delta
f\{y(t_{\text{\tiny f}}),y_0\}}$, where the free energy change
$\Delta f\{y(t_{\text{\sc f}}),y_0\} = f\{\beta,y(t_{\text{\sc
f}})\} - f(\beta,y_0)$; and the average work turns out to be
$\<W\>_{\scriptscriptstyle{P}} = \int_{-\infty}^{\infty} dW\, W\,
P(W) = (\hbar/2)\, \{Q^*\,y(t_{\text{\sc f}}) - y_0\}\,
\coth(\beta\hbar y_0/2) \ne e\{\beta,y(t_{\text{\sc f}})\} -
e(\beta,y_0)$ \cite{DEF08}, where $Q^*\{y_0,y(t_{\text{\sc f}})\} =$
\begin{equation}
    \frac{1}{2\,y_0\,y(t_{\text{\sc f}})} \left[y_0^2\, \{y^2(t_{\text{\sc f}})\cdot X^2 + \dot{X}^2\} +
    \{y^2(t_{\text{\sc f}})\cdot Y^2 +
    \dot{Y}^2\}\right]\label{eq:q-fkt1-0}
\end{equation}
with both $X$ and $Y$ expressed in terms of the Airy functions
$\mbox{Ai}$ and $\mbox{Bi}$. We stress that this average work
evaluated with the distribution $P(W)$ is not a quantum-mechanical
expectation value of an observable \cite{TAL07}.

%%%%%%%%%%%%%%%%%%%%%%%%%%%%%%%%%%%%%%%%%%%%%%%%%%%%%%%%%%%%%%%%%%%%%%%%%%%%%
\section{Discussion of the second law within the Drude-Ullersma model}\label{sec:2nd_law_drude}
%%%%%%%%%%%%%%%%%%%%%%%%%%%%%%%%%%%%%%%%%%%%%%%%%%%%%%%%%%%%%%%%%%%%%%%%%%%%%
Now we intend to discuss the second law within an oscillator coupled
to a bath at an arbitrary strength. We do this in the Drude-Ullersma
model for the damping kernel in (\ref{eq:chi_tilde1}) with
$\tilde{\gamma}_d(\omega + i 0^+) = \gamma_{\mbox{\tiny
o}}\,\omega_d/(\omega_d - i \omega)$, where a damping parameter
$\gamma_{\mbox{\tiny o}}$, representing the system-bath coupling
strength, and a cut-off frequency $\omega_d$ \cite{ULL66}. It is
convenient to adopt, in place of $\{y(t), \omega_d,
\gamma_{\mbox{\tiny o}}\}$, the parameters $\{\bar{y}_t, \Omega_t,
\bar{\gamma}_t\}$ through the relations \cite{FOR06}
\begin{eqnarray}\label{eq:parameter_change0}
    &\displaystyle y^2(t)\; =\; \bar{y}_t^2\; \frac{\Omega_t}{\Omega_t\, +\, \bar{\gamma}_t}\;\; ;\;\;
    \omega_d\; =\; \Omega_t\, +\, \bar{\gamma}_t&\n\\
    &\displaystyle\gamma_{\mbox{\tiny o}}\; =\; \bar{\gamma}_t\; \frac{\Omega_t\,
    (\Omega_t\, +\, \bar{\gamma}_t)\, +\, \bar{y}_t^2}{(\Omega_t\, +\, \bar{\gamma}_t)^2}\,.&
\end{eqnarray}
Then it can be shown that
\begin{eqnarray}\label{eq:useful-relations_1}
    \hspace*{-.2cm}&&\Omega_t + z_{t,1} + z_{t,2}\, =\, \omega_d\;\; ;\;\; \Omega_t\,z_{t,1}\,z_{t,2}\, =\, y^2(t)\cdot\omega_d\\
    \hspace*{-.2cm}&&\Omega_t\,z_{t,1} + z_{t,1}\,z_{t,2} + z_{t,2}\,\Omega_t\, =\, \bar{y}_t^2 + \Omega_t\,\bar{\gamma}_t\, =\, y^2(t) +
    \omega_d\,\gamma_{\mbox{\tiny o}}\,,\n
\end{eqnarray}
where $(z_{t,1}, z_{t,2}) = (\bar{\gamma}_t/2 + i \bar{{\mathbf
w}}_t, \bar{\gamma}_t/2 - i \bar{{\mathbf w}}_t)$ with
$\bar{{\mathbf w}}_t = \{\bar{y}_t^2 -
(\bar{\gamma}_t/2)^2\}^{1/2}$. Assuming that $\gamma_{\mbox{\tiny
o}} \ne 0$, the susceptibility (\ref{eq:chi_tilde1}) reduces to
\cite{FOR06,KIM06,KIM07}
\begin{equation}\label{eq:susceptibility_drude1}
    \tilde{\chi}_t^{(d)}(\omega + i 0^+)\; =\; -\frac{1}{M}\,
    \frac{\omega\, +\, i\,(\Omega_t\, +\, z_{t,1}\, +\, z_{t,2})}{(\omega\, +\, i \Omega_t)
    (\omega\, +\, i z_{t,1}) (\omega\, +\, i z_{t,2})}\,.
\end{equation}
In comparison, it turns out that in an isolated case
($\gamma_{\mbox{\tiny o}} = 0$), we have $\bar{\gamma}_t = 0$ and so
$y(t) = \bar{y}_t = \bar{{\mathbf w}}_t$ and $\omega_d = \Omega_t$,
as well as $z_{t,1} = i y(t)$ and $z_{t,2} = -i y(t)$, therefore Eq. (\ref{eq:susceptibility_drude1}) reduces to $(M
\{y^2(t) - \omega^2\})^{-1}$, simply real-valued; in this case, the imaginary
number, $\pi i/\{2 M y(t)\}\cdot\delta\{\omega - y(t)\}$ must be added to this real number for the actual susceptibility.

Now we substitute (\ref{eq:susceptibility_drude1}) into
(\ref{eq:energy_finite_temp1-1}), which gives
\begin{equation}\label{eq:free-energy-drude-model1}
    {\mathcal U}_s^*(t)\; =\; \int_0^{\infty} d\omega\; e(\beta,\omega)\cdot{\mathcal
    P}_t^*(\omega)\,,
\end{equation}
where the distribution
\begin{align}
    {\mathcal P}_t^*(\omega)\; =&\; \frac{1}{\pi}\; \mbox{Im}\left\{\frac{d}{d \omega}\, \ln\tilde{\chi}_t(\omega + i 0^+)\right\}\n\\
    =&\; \frac{1}{\pi} \left[\left\{\sum_{l=1}^3 \frac{\underline{\omega_l}(t)}{\omega^2 +
    \underline{\omega_l}^2(t)}\right\}\, -\, \frac{\omega_d}{\omega^2 +
    \omega_d^2}\right]\tag{\ref{eq:free-energy-drude-model1}a}\label{eq:imaginary-susceptibility1}
\end{align}
with $(\underline{\omega_1}, \underline{\omega_2},
\underline{\omega_3}) := (\Omega_t, z_{t,1}, z_{t,2})$, as well as
with the normalization $\int_0^{\infty} d\omega\,{\mathcal
P}_t^*(\omega) = 1$ for $\gamma_{\mbox{\tiny o}} \ne 0$
\cite{ILK13}. With the aid of (\ref{eq:useful-relations_1}), Eq.
(\ref{eq:imaginary-susceptibility1}) can be rewritten as a compact
expression
\begin{equation}\label{eq:imaginary-susceptibility1-1}
    {\mathcal P}_t^*(\omega)\; =\;
    \frac{\omega_d^2\,\gamma_{\mbox{\tiny o}}\cdot g_t^*(\omega)/\pi}{(\omega^2
    + \omega_d^2)\cdot g_t(\omega)}\,,
\end{equation}
where both factors
\begin{align}
    g_t(\omega)\; &=\; (\omega^2 + \Omega_t^2)\, \{\bar{\gamma}_t^2\,\omega^2 + (\omega^2 -
    \bar{y}_t^2)^2\}\tag{\ref{eq:imaginary-susceptibility1-1}a}\label{eq:fkt_g(w)_3}\\
    g_t^*(\omega)\; &=\; 3\,\omega^4\, +\,
    \{\omega_d^2 - \omega_d\,\gamma_{\mbox{\tiny o}} - y^2(t)\}\,\omega^2\, +\, \{\omega_d\cdot y(t)\}^2\,.\tag{\ref{eq:imaginary-susceptibility1-1}b}\label{eq:fkt_g(w)_1}
\end{align}
It is observed that the polynomial $g_t(\omega)$ is non-negative,
while $g_t^*(\omega)$ can be negative and so can ${\mathcal
P}_t^*(\omega)$; the behaviors of ${\mathcal P}_t^*(\omega)$ are
plotted in Fig. \ref{fig:fig1}. Then it is easy to see that in the
classical case, Eq. (\ref{eq:free-energy-drude-model1}) reduces to
${\mathcal U}_{s,\text{\sc cl}}^*(t) = e_{\text{\sc cl}}(\beta)$
regardless of the coupling strength $\gamma_{\mbox{\tiny o}}$.
Likewise, we also have the free energy expressed as
\begin{equation}
    {\mathcal F}_s^*(t)\; =\; \int_0^{\infty} d\omega\; f(\beta,\omega)\cdot{\mathcal P}_t^*(\omega)\,.\label{eq:free-energy-reduced-partition1}
\end{equation}
In comparison, Eq. (\ref{eq:imaginary-susceptibility1}) identically
vanishes in an isolated case; however, for the aforementioned
reason, we have ${\mathcal P}_t^*(\omega) = \delta\{\omega - y(t)\}$
indeed in this case. Therefore, we may say that the smoothness of
the distribution (\ref{eq:imaginary-susceptibility1-1}) reflects the
system-bath coupling.

To observe explicitly how the system-bath coupling strength affects
both energies ${\mathcal U}_s^*(t)$ and ${\mathcal F}_s^*(t)$, we
apply to (\ref{eq:imaginary-susceptibility1}) the identity obtained
from the interplay between generalized functions and the theory of
moments \cite{KAN04} such that
\begin{equation}\label{eq:g_delta_representation1}
    {\mathcal P}_t^*(\omega)\; =\; \sum_{n=0}^{\infty}
    \frac{(-1)^n\cdot\mu_n^*\cdot\delta^{(n)}\{\omega - y(t)\}}{n!}\,.
\end{equation}
Here the symbol $\delta^{(n)}\{\cdots\}$ denotes the $n$th
derivative, and the $n$th moment $\mu_n^* = \int_{-\infty}^{\infty}
d\omega\, \{\omega - y(t)\}^n\, {\mathcal P}_t^*(\omega)\,
\Theta(\omega)$ with the Heaviside step function $\Theta(\omega)$.
Then, with the aid of (\ref{eq:useful-relations_1}), the formal
decomposition (\ref{eq:g_delta_representation1}) can explicitly be
evaluated as
\begin{eqnarray}\label{eq:probability0-11}
    {\mathcal P}_t^*(\omega) &=& \delta\{\omega - y(t)\}\, -\, \mu_1^*(t)\cdot\delta^{(1)}\{\omega - y(t)\}\, +\\
    && \frac{\mu_2^*(t)}{2}\cdot\delta^{(2)}\{\omega - y(t)\}\, +\,\cdots\,,\n
\end{eqnarray}
where the moments are given by $\mu_0^*(t) = 1$ and
\begin{align}
    \mu_1^*(t) =&\; -y(t)\; +\tag{\ref{eq:probability0-11}a}\label{eq:prob_distribution_delta_sum2}\\
    &\; \frac{1}{\pi}
    \left[\omega_d\, \ln\left\{\frac{\omega_d}{y(t)}\right\}\, -\, \sum_{l=1}^3
    \underline{\omega_l}(t)\cdot
    \ln\left\{\frac{\underline{\omega_l}(t)}{y(t)}\right\}\right]\; ;\n\\
    \mu_2^*(t) =&\; \omega_d\,\gamma_{\mbox{\tiny o}} -
    2\, y(t)\cdot\mu_1^*(t)\; ;\; \mu_3^*(t)\,=\,\cdots\,.\tag{\ref{eq:probability0-11}b}\label{eq:prob_distribution_delta_sum2-1}
\end{align}
It is easy to verify that in the weak-coupling limit
$\gamma_{\mbox{\tiny o}} \to 0$, all moments $\mu_n^* \to 0$ where
$n \geq 1$. The substitution of (\ref{eq:probability0-11}) into
(\ref{eq:free-energy-drude-model1}) then allows us to have
\begin{equation}\label{eq:free-energy-drude-model1-10}
    {\mathcal U}_s^*(t)\, =\, e\{\beta,y(t)\}\, +\, \sum_{n=1}^{\infty} \mu_n^*(t)\cdot\{\partial_{\omega}^n\, e(\beta,\omega)\}_{\omega\to
    y(t)}\,,
\end{equation}
where the summation on the right-hand side reflects all system-bath
coupling. An expression with the same structure immediately follows
for the free energy ${\mathcal F}_s^*(t)$, too.

It is a noteworthy fact that as a simple case, we have
$f(\infty,\omega) = \hbar\omega/2$ at zero temperature, and so Eq.
(\ref{eq:free-energy-reduced-partition1}) can be rewritten as a
reduced expression $\int_0^{\infty} d\omega\, f(\infty,\omega)\cdot
\delta\{\omega - y_0^*(t)\}$ with $y_0^*(t) := \sum_{k=0}
\bar{\omega}_{k,t} - \sum_{j=1} \omega_j$ [cf.
(\ref{eq:energy_finite_temp1-1-1})]. Therefore, it looks like a free
energy of the Hamiltonian $\hat{{\mathcal H}}_s^*(t)$ which is in an
isolated pure state; there is no heat flow between system and bath
at zero temperature. On the other hand, the system-of-interest
$\hat{H}_s(t)$ is in a mixed state due to the system-bath coupling
even at zero temperature [cf. (\ref{eq:density_operator1}) and
(\ref{eq:density_operator10})]. This also suggests to us that the
free energy ${\mathcal F}_s^*(t)$ cannot exactly be associated with
the reduced system $\hat{H}_s(t)$ alone.

Similarly to (\ref{eq:free-energy-drude-model1}), we next introduce another distribution which is
useful for describing the the internal energy $U_s(t)$ beyond the
weak-coupling. By using (\ref{eq:energy1}) with $y_0 \to y(t)$,
we can easily get
\begin{equation}\label{eq:internal-energy1}
    U_s(t)\; =\; \int_0^{\infty} d\omega\; e(\beta,\omega)\cdot
    P_t(\omega)\,,
\end{equation}
where the distribution
\begin{align}\tag{\ref{eq:internal-energy1}a}\label{eq:distribution-fkt1}
    P_t(\omega)\; =\; \frac{M}{\pi}\,\frac{\omega^2 +
    y^2(t)}{\omega}\cdot\mbox{Im}\{\tilde{\chi}_t(\omega + i
    0^+)\}\,,
\end{align}
reducing to $\delta\{\omega - y(t)\}$ in the identically vanishing
coupling. Within the Drude-Ullersma model, we have \cite{KIM07}
\begin{equation}\label{eq:im_chi1}
    \mbox{Im}\{\tilde{\chi}_t(\omega + i 0^+)\}\; =\; -\frac{1}{M}\,\sum_{l=1}^3\,\lambda_t^{(l)}\, \frac{\omega}{\omega^2 +
    \underline{\omega_l}^2(t)}\,,
\end{equation}
directly obtained from (\ref{eq:susceptibility_drude1}). Here the
coefficients
\begin{align}
    &\lambda_t^{(1)}\; =\; \frac{z_{t,1}\,+\,z_{t,2}}{(\Omega_t\,-\,z_{t,1})
    (z_{t,2}\,-\,\Omega_t)}\n\\
    &\lambda_t^{(2)}\; =\; \frac{\Omega_t\,+\,z_{t,2}}{(z_{t,1}\,-\,\Omega_t) (z_{t,2}\,-\,z_{t,1})}\tag{\ref{eq:im_chi1}a}\label{eq:coefficients}\\
    &\lambda_t^{(3)}\; =\; \frac{\Omega_t\,+\,z_{t,1}}{(z_{t,2}\,-\,\Omega_t)
    (z_{t,1}\,-\,z_{t,2})}\n
\end{align}
satisfy the relations
\begin{align}
    &\sum_{l=1}^3\,\frac{\lambda_t^{(l)}}{\underline{\omega_l(t)}}\, =\, -\frac{1}{y^2(t)}\; ;\;
    \sum_{l=1}^3\,\lambda_t^{(l)}\, =\, 0\tag{\ref{eq:im_chi1}b}\label{eq:drude_coefficient_relations}\\
    &\sum_{l=3}^3\,\lambda_t^{(l)}\cdot\underline{\omega_l}(t)\, =\, 1\; ;\; \sum_{l=1}^3\,\lambda_t^{(l)}\cdot\underline{\omega_l}^3(t)\, =\, -y^2(t) - \gamma_{\mbox{\tiny o}}\,\omega_d\n\\
    &\sum_{l=1}^3\,\lambda_t^{(l)}\cdot\underline{\omega_l}^2(t)\, =\, 0\; ;\;
    \sum_{l=1}^3\,\lambda_t^{(l)}\cdot\underline{\omega_l}^4(t)\, =\, -\gamma_{\mbox{\tiny o}}\,\omega_d^2\,,\n
\end{align}
which will be useful below. In the weak-coupling limit
$\gamma_{\mbox{\tiny o}} \to 0$, we easily get
\begin{align}\tag{\ref{eq:im_chi1}c}\label{eq:weak-coupling_coefficients1}
    \left(\lambda_t^{(1)}, \lambda_t^{(2)}, \lambda_t^{(3)}\right) \to \left(0, \frac{1}{2 i\,y(t)}, \frac{-1}{2 i\,y(t)}\right)\,.
\end{align}
Eqs. (\ref{eq:x_correlation1}) and (\ref{eq:x_dot_correlation1}) can
then be expressed explicitly as \cite{KIM07}
\begin{eqnarray}
    \<\hat{q}^2\>_{\beta}(t) &=& \frac{1}{M\,\beta\,y^2(t)} + \frac{\hbar}{\pi M}
    \sum_{l=1}^3 \lambda_t^{(l)}\cdot\psi\left(1 + \frac{\beta \hbar \underline{\omega_l}(t)}{2 \pi}\right)\n\\
    \<\dot{\hat{q}}^2\>_{\beta}(t) &=& \frac{1}{M\,\beta} - \frac{\hbar}{\pi
    M}\, \times\n\\
    && \sum_{l=1}^3 \lambda_t^{(l)}\,\underline{\omega_l}^2(t)\cdot\psi\left(1 + \frac{\beta \hbar \underline{\omega_l}(t)}{2 \pi}\right)\label{eq:x_x_dot_drude_overdamped1}
\end{eqnarray}
in terms of the digamma function $\psi(\cdots)$, respectively, thus
immediately giving the internal energy $U_s(t)$ in its closed form.
Here we also used the relation $\psi(1 + z) = \psi(z) + 1/z$
\cite{ABS74}. From this, we can easily verify that in the classical
limit $\hbar \to 0$, the internal energy $U_s(t)\to e_{\text{\sc
cl}}(\beta)$ regardless of the coupling strength
$\gamma_{\mbox{\tiny o}}$.

Now we substitute (\ref{eq:im_chi1}) with
(\ref{eq:drude_coefficient_relations}) into
(\ref{eq:distribution-fkt1}), which will yield
%\begin{subequations}
\begin{eqnarray}
    P_t(\omega) &=& -\frac{\{\omega^2 + y^2(t)\}}{\pi} \sum_{l=1}^3\,
    \frac{\lambda_t^{(l)}}{\omega^2 + \underline{\omega_l}^2(t)}\n\label{eq:distribution-fkt1-11}\\
    &=& (\omega_d^2\, \gamma_{\mbox{\tiny o}}/\pi)\, \{\omega^2 + y^2(t)\}/g_t(\omega)\,.\label{eq:distribution-fkt1-12}
\end{eqnarray}
%\end{subequations}
Therefore, the distribution $P_t(\omega)$ is non-negative. It is
easy to verify the normalization $\int_0^{\infty} d\omega
P_t(\omega) = 1$ for $\gamma_{\mbox{\tiny o}} \ne 0$. The behaviors
of $P_t(\omega)$ are displayed in Fig. \ref{fig:fig2}. In
comparison, Eq. (\ref{eq:distribution-fkt1-11}) identically vanishes
in an isolated case; however, likewise with ${\mathcal
P}_t^*(\omega)$, we have $P_t(\omega) = \delta\{\omega - y(t)\}$
indeed in this case. We can also obtain, as the counterpart to
(\ref{eq:probability0-11}),
\begin{eqnarray}\label{eq:probability_1}
    P_t(\omega) &=& \delta\{\omega - y(t)\}\, -\,
    \mu_1(t)\cdot\delta^{(1)}\{\omega - y(t)\}\, +\n\\
    && \frac{\mu_2(t)}{2}\cdot\delta^{(2)}\{\omega - y(t)\}\,
    +\,\cdots\,,
\end{eqnarray}
where the moments are given by $\mu_0(t) = 1$ and
\begin{align}
    \mu_1(t) =&\; -y(t)\; +\tag{\ref{eq:probability_1}a}\label{eq:temp_frequency1}\\
    &\; \frac{1}{\pi}\,\sum_{l=1}^3\,\lambda_t^{(l)}\,
    \left\{y^2(t) - \underline{\omega_l}^2(t)\right\}\cdot\ln\left\{\frac{\underline{\omega_l}(t)}{y(t)}\right\}\n\\
    \mu_2(t) =&\; \omega_d\,\gamma_{\mbox{\tiny o}}/2 -
    2\, y(t)\cdot\mu_1(t)\; ;\; \mu_3(t)\,=\,\cdots\,.\tag{\ref{eq:probability_1}b}\label{eq:temp_frequency1-1}
\end{align}
In the weak-coupling limit $\gamma_{\mbox{\tiny o}} \to 0$, all
moments $\mu_n \to 0$ where $n \geq 1$. The substitution of
(\ref{eq:probability_1}) into (\ref{eq:internal-energy1}) can
immediately give rise to the sum rule for $U_s(t)$, which is the
counterpart to (\ref{eq:free-energy-drude-model1-10}). In addition,
we stress that the probability density $P_t(\omega)$ is not a
quantum-mechanical quantity.

Next we intend to express the distribution ${\mathcal
P}_t^*(\omega)$ in terms of $P_t(\omega)$, which will enable us to
relate the thermodynamic energy ${\mathcal U}_s^*(t)$ directly to
thermodynamic quantities of the reduced system $\hat{H}_s(t)$. We
first compare (\ref{eq:imaginary-susceptibility1-1}) and
(\ref{eq:distribution-fkt1-12}), which easily leads to ${\mathcal
P}_t^*(\omega) = P_t(\omega) + \tilde{\mathcal P}_t^*(\omega)$ with
\begin{eqnarray}\label{eq:final_12}
    \tilde{\mathcal P}_t^*(\omega) &:=& \frac{-2/\pi}{(\omega^2 + \omega_d^2)} \sum_{l=1}^3
    \frac{\lambda_t^{(l)}\cdot\underline{\omega_l}^2(t)}{\omega^2 + \underline{\omega_l}^2(t)}\, \times\n\\
    && \{y^2(t) + \underline{\omega_l}^2(t) + \omega_d\,\gamma_{\mbox{\tiny o}}/2\}\,.
\end{eqnarray}
It is also straightforward to verify that $\int_0^{\infty}
\tilde{\mathcal P}_t^*(\omega)\,d\omega = 0$. In the Ohmic limit,
Eq. (\ref{eq:final_12}) vanishes. Then we can get
\begin{equation}\label{eq:imaginary-susceptibility4}
    {\mathcal P}_t^*(\omega)\; =\;
    \left[1 + \frac{\{\omega^2 - y^2(t)\}^2}{\omega_d\,\gamma_{\mbox{\tiny o}}\,\{\omega^2 + y^2(t)\}}\right]\, P_t(\omega)\,
    -\, \frac{\omega_d/\pi}{\omega^2 + \omega_d^2}\,.
\end{equation}
By substituting this into (\ref{eq:free-energy-drude-model1}) and
then applying Eqs. (\ref{eq:dissipation_fluctuation_theorem}) and
(\ref{eq:internal-energy1}), we can finally arrive at the expression
\begin{equation}\label{eq:energy-drude-model5}
    {\mathcal U}_s^*(t)\; =\; U_s(t)\, +\, \tilde{\mathcal U}_s^*(t)\,,
\end{equation}
where the coupling-induced term (cf. Appendix \ref{sec:appendix1})
\begin{align}
    & \tilde{\mathcal U}_s^*(t)\, =\, \int_0^{\infty} d\omega\; e(\beta,\omega)\cdot\tilde{\mathcal
    P}_t^*(\omega)\tag{\ref{eq:energy-drude-model5}a}\label{eq:energy-drude-model5_03}\\
    =\; &
    \frac{\hbar \omega_d}{2 \pi}\cdot\psi\left(1 + \frac{\beta \hbar \omega_d}{2 \pi}\right)\, +\,
    \frac{\hbar}{2 \pi\,\omega_d\,\gamma_{\mbox{\tiny o}}}\, \times\tag{\ref{eq:energy-drude-model5}b}\label{eq:energy-drude5-03-1}\\
    \;& \sum_{l=1}^3
    \lambda_t^{(l)}\,\{y^2(t) + \underline{\omega_l}^2(t)\}^2\cdot\psi\left(1 + \frac{\beta \hbar \underline{\omega_l}(t)}{2 \pi}\right)\,.\n
\end{align}
It can be shown that this vanishes indeed with $\gamma_{\mbox{\tiny
o}} \to 0$. This also vanishes in the classical limit $\hbar \to 0$.
Eq. (\ref{eq:energy-drude5-03-1}) is displayed in Fig.
\ref{fig:fig3}.

Similarly, the free energy can also be expressed as
\begin{equation}\label{eq:free-energy-drude-model5}
    {\mathcal F}_s^*(t)\; =\; \bar{F}_s(t)\, +\, \tilde{\mathcal F}_s^*(t)\,,
\end{equation}
where a generalized ``free energy'' $\bar{F}_s(t) = \int_0^{\infty}
d\omega\,f(\beta,\omega)\cdot P_t(\omega)$, and
\begin{align}
    & \tilde{\mathcal F}_s^*(t)\, =\, \int_0^{\infty} d\omega\, f(\beta,\omega)\cdot\tilde{\mathcal P}_t^*(\omega)\tag{\ref{eq:free-energy-drude-model5}a}\label{eq:final_17}\\
     =\; &
    \frac{1}{\beta}\cdot\ln\left\{\Gamma\left(1 + \frac{\beta \hbar \omega_d}{2 \pi}\right)\right\}\, +\,
    \frac{1}{\beta\,\omega_d\,\gamma_{\mbox{\tiny o}}}\, \times\tag{\ref{eq:free-energy-drude-model5}b}\label{eq:final_17-01}\\
    \;& \sum_{l=1}^3
    \frac{\lambda_t^{(l)}}{\underline{\omega_l}(t)}\,
    \{y^2(t) + \underline{\omega_l}^2(t)\}^2\cdot\ln\left\{\Gamma\left(1 + \frac{\beta \hbar \underline{\omega_l}(t)}{2 \pi}\right)\right\}\,,\n
\end{align}
expressed in terms of the gamma function $\Gamma(\cdots)$. It can be
shown that Eq. (\ref{eq:final_17-01}) vanishes with
$\gamma_{\mbox{\tiny o}} \to 0$, as well as with $\hbar \to 0$. Fig.
\ref{fig:fig4} displays different behaviors of
(\ref{eq:final_17-01}).

Now we consider the free energy change $\Delta{\mathcal
F}_s^*(t_{\text{\sc f}}) = \int_0^{\infty} d\omega\,
f(\beta,\omega)\,\{{\mathcal P}_{t_{\text{\tiny f}}}^*(\omega) -
{\mathcal P}_0^*(\omega)\}$. This turns out to be identical to the
quantum-mechanical average value
\begin{equation}\label{eq:final_1-1}
    {\mathcal W}_{\text{\sc rev}}\{\beta,y(t_{\text{\sc f}})\}\; =\; \int_{y_0}^{y(t_{\text{\tiny f}})} dy\,
    \left\<\frac{\partial \hat{H}_s\{y(t)\}}{\partial
    y}\right\>_{\beta}\,,
\end{equation}
evaluated along an isothermal process, i.e., in the infinitesimally
slow variation of frequency (cf. Appendix \ref{sec:appendix2}).
However, this ``work'' ${\mathcal W}_{\text{\sc
rev}}\{\beta,y(t_{\text{\sc f}})\}$ may conceptually not be
interpreted as a minimum average work performed on the reduced
system $\hat{H}_s(t)$, due to the fact that the actual average work
should be defined as an average value of a classical stochastic
variable with transition probabilities derived from quantum
mechanics, rather than as an expectation value of some ``work''
operator \cite{TAL07}; accordingly, the minimum average work comes
out when the individual work of each run is performed only for the
reversible process.

Next let us discuss the second law of thermodynamics within the
system-of-interest $\hat{H}_s(t)$ in terms of $\Delta
U_s(t_{\text{\sc f}}) = \int_0^{\infty} d\omega\,
e(\beta,\omega)\,\{P_{t_{\text{\tiny f}}}(\omega) - P_0(\omega)\}$
and $\Delta{\mathcal F}_s^*(t_{\text{\sc f}})$. To do so, we
reinterpret an entire process $\{y(t)\,|\,0 \leq t \leq t_{\text{\sc
f}}\}$ as a sum of the following three sub-processes, based on the
fact that all thermodynamic state functions are path-independent; in
the sub-process (I), we completely decouple the system oscillator
$\hat{H}_s(0)$ from a bath by letting the non-vanishing coupling
strengths $c_j \to 0$ in an isothermal fashion. The internal energy
change through this first sub-process is given by $\Delta U_{s,1} =
e(\beta,y_0) - U_s(0)$ while the free energy change needed for
decoupling the system from the bath, $\Delta{\mathcal F}_{s,1}^* =
f(\beta,y_0) - {\mathcal F}_s^*(0)$ \cite{FOR06,KIM06,KIM07}. In the
next sub-process (II) we vary the frequency $y(t)$ of the resultant
isolated oscillator according to the pre-determined protocol,
followed by its coupling weakly to the bath which makes the
oscillator come back to the thermal equilibrium at temperature $T$.
The relevant internal energy change and free energy change are
$\Delta U_{s,2}(t_{\text{\sc f}}) = e\{\beta,y(t_{\text{\sc f}})\} -
e(\beta,y_0)$ and $\Delta{\mathcal F}_{s,2}^*(t_{\text{\sc f}}) =
f\{\beta,y(t_{\text{\sc f}})\} - f(\beta,y_0)$, respectively. In
this sub-process, the JE (\ref{eq:jarzynski1}) holds true. In the
last sub-process (III) we increase the coupling strengths $c_j$ up
to their original values in an isothermal fashion; the internal
energy change $\Delta U_{s,3} = U_s(t_{\text{\sc f}}) -
e\{\beta,y(t_{\text{\sc f}})\}$, and the free energy change needed
for coupling the system to the bath, $\Delta{\mathcal F}_{s,3}^* =
{\mathcal F}_s^*(t_{\text{\sc f}}) - f\{\beta,y(t_{\text{\sc
f}})\}$. As a result, the total internal energy change and free
energy change through sub-processes (I)-(III) equal $\Delta
U_s(t_{\text{\sc f}})$ and $\Delta{\mathcal F}_s^*(t_{\text{\sc
f}})$, respectively.

We remind that the inequality $\Delta U_{s,3} < \Delta{\mathcal
F}_{s,3}^*$ is valid (so is $\Delta U_{s,1} > \Delta{\mathcal
F}_{s,1}^*$), particularly in the low-temperature regime
\cite{KIM07,FOR06,KIM06}. Therefore, this inequality says that if we
consider a next decoupling process after (III) and then identified
the free change $\Delta{\mathcal F}_{s,3}^*$ with the (maximum)
useful work spontaneously releasable from the coupled system
$\hat{H}_s(t_{\text{\sc f}})$, then the second law within the system
$\hat{H}_s(t_{\text{\sc f}})$ could be violated. This tells us that
the free energy change $\Delta{\mathcal F}_s^*(t_{\text{\sc f}})$
cannot be the (minimum) actual work performed on the system
$\hat{H}_s(t)$ beyond the weak-coupling limit, and so the JE
(\ref{eq:haenggi-jarzynski_1-1}) is not allowed to associate itself
directly with the second law within the coupled oscillator.

Comments deserve here. From Eqs. (\ref{eq:internal-energy1}) and
(\ref{eq:energy-drude-model5})-(\ref{eq:free-energy-drude-model5})
it is tempted to rewrite the internal energy as $U_s(t) =
-\partial_{\beta} \ln \bar{Z}_{\beta}\{y(t)\}$ in terms of a
generalized partition function $\bar{Z}_{\beta}\{y(t)\} :=
\exp\{\int_0^{\infty} d\omega\,\ln z_{\beta}(\omega)\,P_t(\omega)\}$
beyond the weak-coupling limit, and the resultant ``free energy'' as
$\bar{F}_s(t) = -\beta^{-1}\,\ln \bar{Z}_{\beta}\{y(t)\}$, with
$\bar{F}_s(t_{\text{\sc f}}) \ne {\mathcal F}_s^*(t_{\text{\sc
f}})$. In the classical case, on the other hand, Eq.
(\ref{eq:final_17}) vanishes, and so $\bar{F}_{s,{\text{\sc
cl}}}(t_{\text{\sc f}}) = {\mathcal F}_{s,{\text{\sc
cl}}}^*(t_{\text{\sc f}})$ indeed, explicitly given by
\begin{equation}\label{eq:final_17-0}
    \bar{F}_{s,{\text{\sc cl}}}(t)\; =\; \frac{1}{2\beta} \sum_{l=1}^3 \frac{\lambda_t^{(l)}\,\{\underline{\omega_l}^2(t) -
    y^2(t)\}\cdot\ln\{\underline{\omega_l}(t)\}}{\underline{\omega_l}(t)}\,.
\end{equation}
It is also notable that Eq. (\ref{eq:final_17-0}) is independent of
$\hbar$, whereas this is not true for $f_{\text{\sc
cl}}(\beta,\omega) = \{\ln(\beta\hbar\omega)\}/\beta$. This means
that the classical free energy can be conceptually rescued only when
the system-bath coupling is reflected (cf. of course, $\Delta
f_{\text{\sc cl}}(\beta,\omega) = (\,\ln\{y(t_{\text{\sc
f}})/y_0\})/\beta$ with $\{\omega\,|\,y_0\to y(t_{\text{\sc f}})\}$,
independent of $\hbar$). This free energy $\bar{F}_s(t)$ will be
employed below for our discussion of a generalized Jarzynski
equality.

%%%%%%%%%%%%%%%%%%%%%%%%%%%%%%%%%%%%%%%%%%%%%%%%%%%%%%%%%%%%%%%%%%%%%%%%%%%%
\section{Quantum Jarzynski equality beyond the weak-coupling limit}\label{sec:Jarzynski}
%%%%%%%%%%%%%%%%%%%%%%%%%%%%%%%%%%%%%%%%%%%%%%%%%%%%%%%%%%%%%%%%%%%%%%%%%%%%%
%
We will introduce a generalized Jarzynski equality consistent with
the second law within the oscillator coupled to a bath at an
arbitrary strength. This will need an appropriate definition of the
work performed on the system. Therefore we first consider a
reversible process in which it is straightforward to evaluate the
work. Then the generalized free energy change in the variation of
frequency can be expressed as
\begin{equation}\label{eq:final17-2}
    \Delta\bar{F}_s(t_{\text{\tiny f}})\; =\; \int_0^{\infty} d\omega\, \Delta f(\omega,y_0)\cdot \{P_{t_{\text{\tiny f}}}(\omega) -
    P_0(\omega)\}\,,
\end{equation}
where $\Delta f(\omega,y_0) = f(\beta,\omega) - f(\beta,y_0)$. From
this, a generalized Jarzynski equality (GJE) beyond the
weak-coupling limit is introduced as
\begin{equation}\label{eq:final_9}
    \Delta\bar{F}_s(t_{\text{\tiny f}}) = -\frac{1}{\beta}
    \int_0^{\infty} d\omega\,\{\ln\<e^{-\beta\,W(\beta,\omega)}\>_{\scriptscriptstyle{P}}\}\,\{P_{t_{\text{\tiny f}}}(\omega) -
    P_0(\omega)\}
\end{equation}
where the average $\<\cdots\>_{\scriptscriptstyle{P}}$ is carried
out with the work distribution $P(W)$ in (\ref{eq:probability_1-1})
for an isolated oscillator in the frequency variation $\{y_0 \to
\omega(t_{\text{\tiny f}})\}$, as well as the probability density
$P_t(\omega)$ reflects the actual coupling between $\hat{H}_s(t)$
and $\hat{H}_b$.

In an isolated case the GJE easily reduces to the known form
(\ref{eq:jarzynski1}). As shown in (\ref{eq:final_9}), we now need
to deal with a sum of the Jarzynski equalities, each being valid for
an isolated system, with the initial and final ``weights'' $P_0$ and
$P_{t_{\text{\tiny f}}}$ obtained directly from the susceptibility
$\chi_t(\omega)$ (cf. Appendix \ref{sec:appendix2}). Technically
this means that we first turn off the coupling $c_j\mbox{'s} \to 0$,
which makes the initial reduced density matrix
(\ref{eq:density_operator1}) reduce to the canonical thermal state
$e^{-\beta \hat{H}_s(0)}/z_{\beta}(y_0)$; next we carry out the JE
processes independently for many different frequencies $\omega$'s
with the two weights. From this, we can extract the free energy
change $\Delta\bar{F}_s(t_{\text{\tiny f}})$, without a measurement
of any other ``work'' directly on the coupled system. It is
instructive to remind that the JE (\ref{eq:haenggi-jarzynski_1-1})
can also be rewritten as
\begin{equation}
    \Delta{\mathcal F}_s^*(t_{\text{\tiny f}}) = -\frac{1}{\beta}\int_0^{\infty} d\omega\,\{\ln\<e^{-\beta\,W(\beta,\omega)}\>\}\,\{{\mathcal P}^*_{t_{\text{\tiny f}}}(\omega) -
    {\mathcal P}^*_0(\omega)\}
\end{equation}
in its form, where the distributions ${\mathcal P}_t^*(\omega)$'s
come from the susceptibility as well [cf.
(\ref{eq:imaginary-susceptibility1}) and
(\ref{eq:free-energy-reduced-partition1})], but not guaranteed to be
non-negative.

Next, in order to observe explicitly the deviation of the GJE from
the JE in its known form, we substitute the sum rule
(\ref{eq:probability_1}) into (\ref{eq:final_9}), which yields
\begin{eqnarray}\label{eq:exact-je-2-1}
    \Delta\bar{F}_s(t_{\text{\tiny f}}) &=& -\frac{1}{\beta} \left[\ln \left\<e^{-\beta\,W\{\beta,y(t_{\text{\tiny f}})\}}\right\>_{\scriptscriptstyle{P}}\, +\right.\\
    && \left.\sum_{n=1}^{\infty} \frac{\mu_n(t)}{n!}\, \left(\frac{\partial}{\partial y}\right)^n \left.\ln \left\<e^{-\beta\,W\{\beta,y(t)\}}\right\>_{\scriptscriptstyle{P}}\right|_{t=0}^{t=t_{\text{\tiny
    f}}}\right]\,.\n
\end{eqnarray}
Therefore we can now look into the sufficiently weak, but not
necessarily vanishingly small, coupling regime, simply by adding the
low-order coupling-induced terms (\ref{eq:temp_frequency1}) and
(\ref{eq:temp_frequency1-1}). On the other hand, beyond the weak
coupling limit we cannot simply neglect higher-order moments (cf.
Fig. \ref{fig:fig5} as well as Figs. 1, 2). As a result, we may say
that the JE (\ref{eq:jarzynski1}) is exactly valid only in the
vanishingly small coupling limit.

Now we briefly discuss (\ref{eq:exact-je-2-1}) in the classical
limit $\beta \hbar \to 0$. Then the term with $n = 1$ easily reduces
to
%\begin{subequations}
\begin{equation}\label{eq:classical-limit1}
    \mu_1(t)\cdot\partial_y f\{\beta,y(t)\}\Big|_{t=0}^{t=t_{\text{\tiny f}}}\, \to\,
    \frac{1}{\beta}\, \left\{\frac{\mu_1(t_{\text{\sc f}})}{y(t_{\text{\sc f}})}\, -\, \frac{\mu_1(0)}{y_0}\right\}\,,
\end{equation}
%\end{subequations}
non-vanishing indeed. Likewise, so are all terms with $n \geq 2$.
Therefore, even the classical JE (\ref{eq:jarzynski1}) does not
exactly hold true any longer beyond the weak-coupling limit.

It is also instructive to add remarks on the effect of system-bath
coupling to the Jarzynski equality (\ref{eq:final_9}); in an
isolated case, albeit the JE in its known form is well-known, we
cannot perform an isothermal process, in which the (minimum) average
work exactly amounts to the free energy change. And in an isothermal
process we might think heat exchange through the system-bath
coupling at every single moment in such a way that we switch off the
coupling (``decoupling'') and then perform the external perturbation
$\{y_0 \to y(t_{\text{\sc f}})\}$ in the resultant isolated case,
followed by contacting with the bath again (``coupling''); so the
total heat exchange over the actual isothermal process would be
equivalent to the amount of final thermal relaxation leading to the
end equilibrium state in the above picture a pair of decoupling and
coupling is added to. However, as discussed after
(\ref{eq:final_1-1}), this picture could lead to a violation of the
second law, and so is not acceptable. Moreover, in order to make the
JE useful, the system under consideration needs to be sufficiently
small-scaled, in which the work fluctuations are observable; in this
scale, however, the system-bath coupling is normally non-negligible.
As a result, we may argue that the usefulness of the JE in its known
form is fairly limited.

Now we discuss the relevance of the GJE to the second law of
thermodynamics within the system-of-interest $\hat{H}_s(t)$ beyond
the weak-coupling limit. We first introduce the average work in an
irreversible process as
%\begin{subequations}
\begin{equation}
    \hspace*{-.2cm}\overline{\<W(t_{\text{\sc f}})\>} = \int_0^{\infty} d\omega\,\left\<W\{\beta,\omega(t_{\text{\sc f}})\}\right\>_{\scriptscriptstyle{P}}\cdot\{P_{t_{\text{\tiny f}}}(\omega) -
    P_0(\omega)\}\,,\label{eq:generalized-average-work-1}
\end{equation}
%\end{subequations}
where each partial average work
$\<W(\beta,\omega)\>_{\scriptscriptstyle{P}}$ is explicitly given by
$(\hbar/2) \{Q^*(y_0,\omega)\cdot\omega - y_0\} \coth(\beta\hbar
y_0/2)$ [cf. (\ref{eq:q-fkt1-0})]. As such, the average work
$\overline{\<W(t_{\text{\sc f}})\>}$, being not a quantum-mechanical
expectation value, can be determined without any measurement of the
work directly on the system coupled to a bath. Applying the Jensen
inequality to (\ref{eq:final_9}), we can then obtain an expression
of the second law of thermodynamics beyond the weak-coupling
\begin{equation}\label{eq:exact-je-5}
    \hspace*{-.2cm}\int_0^{\infty} d\omega\,\left\{\<W(\beta,\omega)\>_{\scriptscriptstyle{P}} - \Delta f(\omega,y_0)\right\}\,\{P_{t_{\text{\tiny f}}}(\omega) -
    P_0(\omega)\} \geq 0\,,
\end{equation}
equivalent to $\overline{\<W(t_{\text{\sc f}})\>} \geq
\bar{F}_s(t_{\text{\sc f}})$. It is now needed to ask if this
equality is valid indeed; Figs. \ref{fig:fig6}-\ref{fig:fig7}
demonstrate its validity for many different sets of parameters
$\{y(t), \omega_d, \gamma_{\mbox{\tiny o}}\}$. Subsequently, the
first law allows us to have the heat $ \overline{\<Q(t_{\text{\sc
f}})\>}\, =\, \Delta U_s(t_{\text{\sc f}})\, -\,
\overline{\<W(t_{\text{\sc f}})\>}$.

Lastly, a couple of short comments deserve here. First, we remind
that our approach was made entirely in the local picture
$\hat{H}_s(t)$, rather than the total picture $\hat{H}(t)$.
Accordingly, there is no room appropriate for detecting system-bath
entanglement directly within our generalized Jarzynski equality,
while it was discussed, on the other hand, within the Jarzynski
equality (\ref{eq:haenggi-jarzynski_1-1}) in \cite{VED10}. Secondly,
in practical terms the number of independent experimental runs
needed for obtaining a sufficiently visible convergence of the JE
(\ref{eq:jarzynski1}) grows exponentially with the system size
\cite{GRO05}, and so the computational cost is high enough. This
cost will be even higher when we deal with a system beyond the
weak-coupling limit, due to the additional averaging needed for the
GJE. Further it was shown \cite{GON13} that a significantly faster
convergence of the JE can be achieved via accelerated adiabatic
control. Even in this scheme the computational cost is expected to
increase beyond the weak-coupling limit, from our result.

%%%%%%%%%%%%%%%%%%%%%%%%%%%%%%%%%%%%%%%%%%%%%%%%%%%%%%%%%%%%%%%%%%%%%%%%%%%%%
\section{Concluding remarks}\label{sec:conclusions}
%%%%%%%%%%%%%%%%%%%%%%%%%%%%%%%%%%%%%%%%%%%%%%%%%%%%%%%%%%%%%%%%%%%%%%%%%%%%%
In summary, we derived a generalized Jarzynski equality in the
scheme of a time-dependent quantum Brownian oscillator within the
Drude-Ullersma damping model. This equality is associated with the
second law of thermodynamics (in its generalized form) within the
system oscillator coupled to a bath at an arbitrary strength. This
finding also enables us to look systematically into the coupling
effect on the non-equilibrium thermodynamics of the local
system-of-interest beyond the weak-coupling limit. As a result, the
Jarzynski equality in its original form (and all other relevant
fluctuation theorems) was shown to be valid only in the vanishingly
small coupling limit, which fact also holds true in the classical
limit of $\beta\hbar \to 0$.

We believe that our finding will provide a useful starting point for
derivation of a generalized Jarzynski equality associated with the
second law in more generic quantum dissipative systems. In fact, if
a smooth probability density, such as $P_t(\omega)$ in
(\ref{eq:internal-energy1}), reflecting the system-bath coupling is
explicitly available, this derivation becomes conceptually rather a
straightforward issue, while the technical procedure for an exact
evaluation of such a probability density would be a formidable task
for a broad class of nonlinear systems.

%
%%%%%%%%%%%%%%%%%%%%%%%%%%%%%%%%%%%%%%%%%%%%%%%%%%%%%%%%%%%%%%%%%%%
\section*{Acknowledgments}
%%%%%%%%%%%%%%%%%%%%%%%%%%%%%%%%%%%%%%%%%%%%%%%%%%%%%%%%%%%%%%%%%%%
The author thanks G. Mahler (Stuttgart), G.J. Iafrate (NC State),
and J. Kim (KIAS) for helpful remarks. He also acknowledges
financial support provided by the U.S. Army Research Office (Grant
No. W911NF-13-1-0323).

%%%%%%%%%%%%%%%%%%%%%%%%%%%%%%%%%%%%%%%%%%%%%%%%%%%%%%%%%%%%%%%%%%%%%%%%%%%%%
\appendix\section{: Detailed derivation of Eq.
(\ref{eq:energy-drude5-03-1})} \label{sec:appendix1}
%%%%%%%%%%%%%%%%%%%%%%%%%%%%%%%%%%%%%%%%%%%%%%%%%%%%%%%%%%%%%%%%%%%%%%%%%%%%%
We first substitute the identity \cite{ING98}
\begin{equation}\label{eq:hyperbolic-cotangent1}
    e(\beta,\omega)\, =\, \frac{\hbar \omega}{2}\,\coth\left(\frac{\beta \hbar \omega}{2}\right)\, =\,
    \frac{1}{\beta}\,\left(1 + 2 \sum_{n=1}^{\infty} \frac{\omega^2}{\omega^2 +
    \nu_n^2}\right)\,,
\end{equation}
with $\nu_n = 2\pi n/(\beta \hbar)$, into
(\ref{eq:energy-drude-model5_03}), finally leading to
$\tilde{\mathcal U}_s^*(t) =$
\begin{equation}\label{eq:final_13}
    \frac{-\omega_d\,\gamma_{\mbox{\tiny o}}}{\beta} \sum_{n=1}^{\infty}
    \frac{\nu_n^2}{(\nu_n + \omega_d)(\nu_n + \Omega_t)(\nu_n + z_{1,t})(\nu_n + z_{2,t})}\,.
\end{equation}
This can be rewritten in terms of the digamma function as
\begin{eqnarray}\label{eq:final_14}
    \hspace*{-.1cm}&&\tilde{\mathcal U}_s^*(t)\, =\,
    \frac{\hbar \omega_d \gamma_{\mbox{\tiny o}}}{2\pi}\, \times\\
    \hspace*{-.1cm}&&\sum_{j=0}^3 \frac{\psi\{1 +
    \beta \hbar \underline{\omega_j}(t)/2\pi\}\cdot\underline{\omega_j}^2(t)}{\{\underline{\omega_j}(t) -
    \underline{\omega_{j+1}}(t)\}\{\underline{\omega_j}(t) - \underline{\omega_{j+2}}(t)\}\{\underline{\omega_j}(t) -
    \underline{\omega_{j+3}}(t)\}}\n
\end{eqnarray}
where $\underline{\omega_0} := \omega_d$, and the subscript
$\underline{j} = j\,\mbox{(mod 4)}$. Subsequently, with the aid of
(\ref{eq:drude_coefficient_relations}) and
(\ref{eq:x_x_dot_drude_overdamped1}) we can finally arrive at the
expression in (\ref{eq:energy-drude5-03-1}).

%%%%%%%%%%%%%%%%%%%%%%%%%%%%%%%%%%%%%%%%%%%%%%%%%%%%%%%%%%%%%%%%%%%%%%%%%%%%%
\section{: Evaluation of Eq. (\ref{eq:final_1-1})}
\label{sec:appendix2}
%%%%%%%%%%%%%%%%%%%%%%%%%%%%%%%%%%%%%%%%%%%%%%%%%%%%%%%%%%%%%%%%%%%%%%%%%%%%%
We begin with [cf. (\ref{eq:total_hamiltonian2-1})]
\begin{equation}\label{eq:work_def1}
    \left\<\frac{\partial \hat{H}_s\{y(t)\}}{\partial y}\right\>_{\beta}\, =\,
    M\,y\,\left\{\<\hat{q}^2\>_{\beta}(y)\right\}
\end{equation}
at every single frequency value $y(t)$. With the aid of
(\ref{eq:x_correlation1}) and $e(\beta,\omega) =
\omega\,\partial_{\omega} f(\beta,\omega)$, it turns out that
\begin{eqnarray}\label{eq:final_1}
    {\mathcal W}_s\{\beta,y(t_{\text{\sc f}})\} &=& \frac{2M}{\pi} \int_0^{\infty} d\omega\, (\partial_{\omega} f)\, \times\n\\
    && \int_{y_0}^{y(t_{\text{\tiny f}})} dy\; y\; \mbox{Im}\{\tilde{\chi}_t(\omega + i\,0^+)\}\,.
\end{eqnarray}
Next, with the aid of (\ref{eq:useful-relations_1}) we can express
the susceptibility (\ref{eq:im_chi1}) in terms of $\{y(t), \omega_d,
\gamma_{\mbox{\tiny o}}\}$ as $\mbox{Im}\{\tilde{\chi}_t(\omega + i
0^+)\} = \omega_d^2\,\gamma_{\mbox{\tiny
o}}\,\omega/\{M\,\Upsilon_t(\omega)\}$, where
\begin{eqnarray}\label{eq:final_2}
    &&\Upsilon_t(\omega)\, =\, \omega^6 + \{\omega_d\,(\omega_d - 2 \gamma_{\mbox{\tiny o}}) - 2\,y^2(t)\}\,\omega^4 +\\
    &&\{(\omega_d\,\gamma_{\mbox{\tiny o}})^2 - 2\,\omega_d\,(\omega_d - \gamma_{\mbox{\tiny o}})\,y^2(t) + y^4(t)\}\,\omega^2 + \omega_d\cdot y^2(t)\,.\n
\end{eqnarray}
Substituting this into (\ref{eq:final_1}), we can finally obtain
\begin{eqnarray}\label{eq:final_4}
    {\mathcal W}_s\{\beta,y(t_{\text{\sc f}})\} &=& \frac{\omega_d^2\,\gamma_{\mbox{\tiny o}}}{\pi} \int_0^{\infty}
    \frac{d\omega\, (\omega\,\partial_{\omega} f)}{\omega^2 +
    \omega_d^2}\, \times\n\\
    && \int_{y_0^2}^{y^2(t_{\text{\tiny f}})} \frac{dz}{z^2 + b z + c}\,,
\end{eqnarray}
where $z := y^2(t)$, and
\begin{align}
    b\; :=&\; \frac{2\,\omega^2\,(\omega_d\,\gamma_{\mbox{\tiny o}} - \omega_d^2 - \omega^2)}{\omega^2 + \omega_d^2}\tag{\ref{eq:final_4}a}\label{eq:final_5}\\
    c\; :=&\; \frac{\omega^2\,\{\omega^4 + (\omega_d^2 - 2\,\omega_d\,\gamma_{\mbox{\tiny o}})\,\omega^2 + (\omega_d\,\gamma_{\mbox{\tiny o}})^2\}}{\omega^2 +
    \omega_d^2}\,.\n
\end{align}
Using the relation \cite{GRA07}
\begin{equation}\label{eq:final_40}
    \int \frac{dz}{z^2 + b z + c}\; =\; \frac{-2}{(b^2 -
    4c)^{1/2}}\cdot\mbox{arctanh}\left\{\frac{2z + b}{(b^2 - 4c)^{1/2}}\right\}
\end{equation}
where $\mbox{arctanh}(z) = \{\ln(1+z) - \ln(1-z)\}/2$, we can arrive
at the expression
\begin{eqnarray}\label{eq;final_6}
    &&{\mathcal W}_s\{\beta,y(t_{\text{\sc f}})\}\, =\, \frac{1}{\pi} \int_0^{\infty} d\omega\, (\partial_{\omega} f)\, \times\\
    &&\left.\mbox{arctan}\left\{\frac{(\omega^2 + \omega_d^2)\cdot y^2(t) - \omega^2\,\omega_d\,(\omega_d - \gamma_{\mbox{\tiny o}}) -
    \omega^4}{\omega_d^2\,\gamma_{\mbox{\tiny o}}\,\omega}\right\}\right|_{t=0}^{t=t_{\text{\tiny f}}}\n
\end{eqnarray}
(note that $\gamma_{\mbox{\tiny o}} \not\equiv 0$). By integration
by parts, this can be transformed into
\begin{equation}\label{eq:final_8}
    {\mathcal W}_s\{\beta,y(t_{\text{\sc f}})\}\; =\; \int_0^{\infty} d\omega\,
    f(\beta,\omega)\cdot \{\bar{\mathcal P}_{t_{\text{\tiny f}}}^*(\omega) - \bar{\mathcal P}_0^*(\omega)\}\,,
\end{equation}
where the distribution $\bar{\mathcal P}_t(\omega) :=$
\begin{align}
    &\frac{1}{\pi}\,\frac{d}{d\omega}\,\mbox{arctan}\left\{\frac{\omega^4 +
    \omega_d\,(\omega_d - \gamma_{\mbox{\tiny o}})\,\omega^2 - (\omega^2 + \omega_d^2)\cdot y^2(t)}{\omega_d^2\,\gamma_{\mbox{\tiny
    o}}\,\omega}\right\}\tag{\ref{eq:final_8}a}\label{eq:final_80}\\
    &=\; (\omega_d^2\, \gamma_{\mbox{\tiny o}}/\pi)\cdot
    g_t^*(\omega)/\{g_t(\omega)\cdot(\omega^2 + \omega_d^2)\}\tag{\ref{eq:final_8}b}\; =\; {\mathcal P}_t^*(\omega)\,.\label{eq:final_81}
\end{align}
Here we also used both $(d/dx)\,\mbox{arctan}(x) = 1/(1 + x^2)$ and
the fact that Eq. (\ref{eq:fkt_g(w)_3}) can be rewritten in terms of
$\{y(t), \omega_d, \gamma_{\mbox{\tiny o}}\}$ as
\begin{eqnarray}
    \hspace*{-.2cm}g_t(\omega) &=& \omega^6\, +\,
    \{\omega_d^2 - 2\,\omega_d\,\gamma_{\mbox{\tiny o}} - 2\,y^2(t)\}\,\omega^4\, +\, \{\omega_d\cdot y^2(t)\}^2\n\\
    \hspace*{-.2cm}&& +\, \left[\{\omega_d\,\gamma_{\mbox{\tiny o}} + y^2(t)\}^2 - 2\,\{\omega_d\cdot y(t)\}^2\right]\,\omega^2\,.\label{eq:fkt_g(w)_2}
\end{eqnarray}

\newpage
\onecolumn{
\begin{figure}[htb]
\centering\hspace*{-2.5cm}{
\includegraphics[scale=0.75]{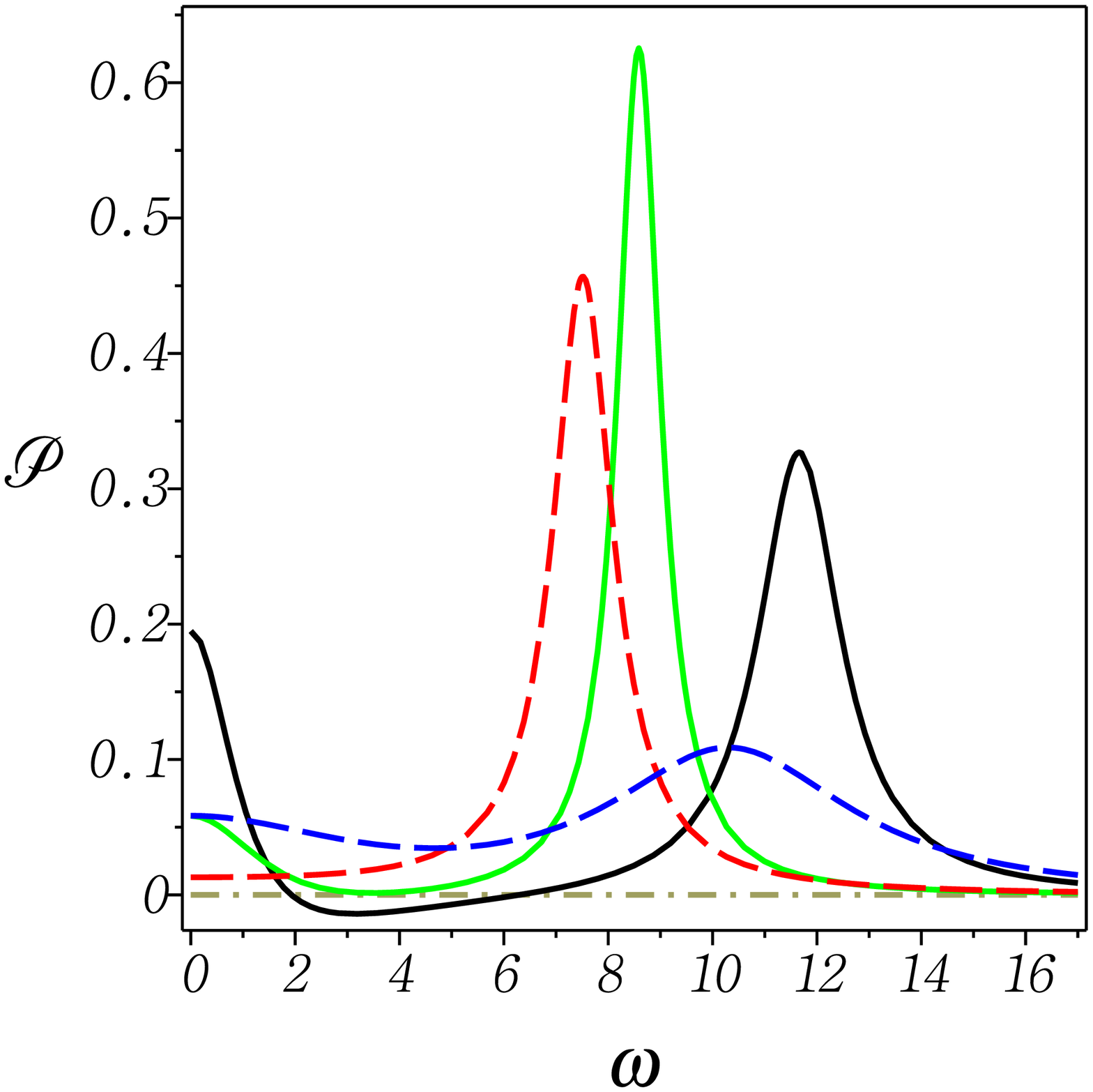}
\caption{\label{fig:fig1}}}
\end{figure}
Fig.~\ref{fig:fig1}: (Color online) The distribution ${\mathcal P} =
{\mathcal P}_0^*(\omega)$ given in
(\ref{eq:imaginary-susceptibility1-1}). Here we set $y_0 = 7$. (I)
Solid lines with $\omega_d = 3$, from top to bottom in the maximum
values, 1st) green: $\gamma_{\mbox{\tiny o}} = 9$; 2nd) black:
$\gamma_{\mbox{\tiny o}} = 30$ (can be negative). (II) Dash lines
with $\omega_d = 10$, likewise, 1st) red: $\gamma_{\mbox{\tiny o}} =
2$; 2nd) blue: $\gamma_{\mbox{\tiny o}} = 9$, in comparison with
${\mathcal P} \equiv 0$ (khaki dashdot); cf. ${\mathcal P} \to
\delta(\omega - 7)$ with $\gamma_{\mbox{\tiny o}} \to 0$.
\newpage
\begin{figure}[htb]
\centering\hspace*{-2.5cm}{
\includegraphics[scale=0.75]{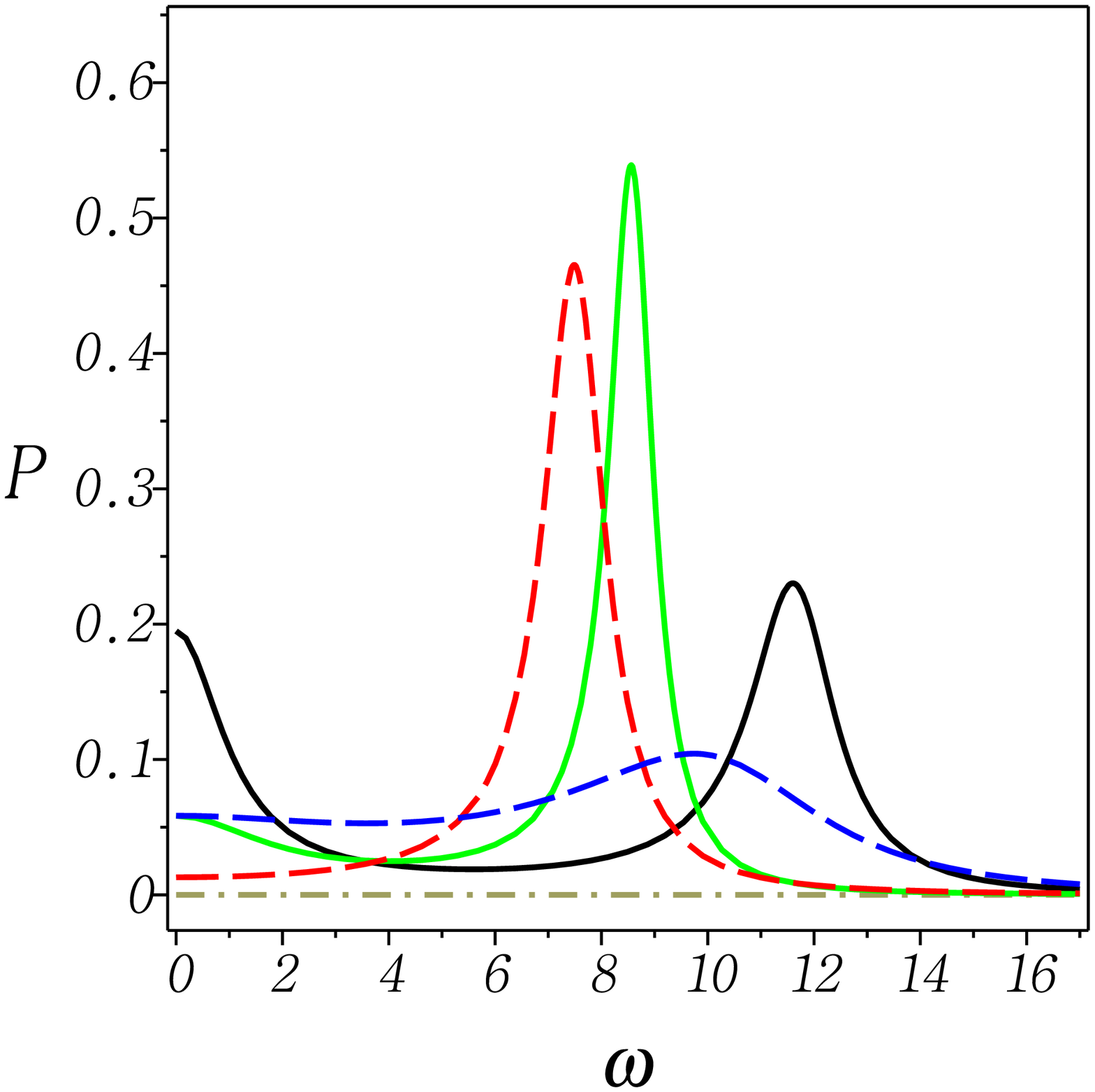}
\caption{\label{fig:fig2}}}
\end{figure}
Fig.~\ref{fig:fig2}: (Color online) The distribution $P =
P_0(\omega)$ given in (\ref{eq:distribution-fkt1-11}). The same as
in Fig. \ref{fig:fig1}.
\newpage
\begin{figure}[htb]
\centering\hspace*{-2.5cm}{
\includegraphics[scale=0.75]{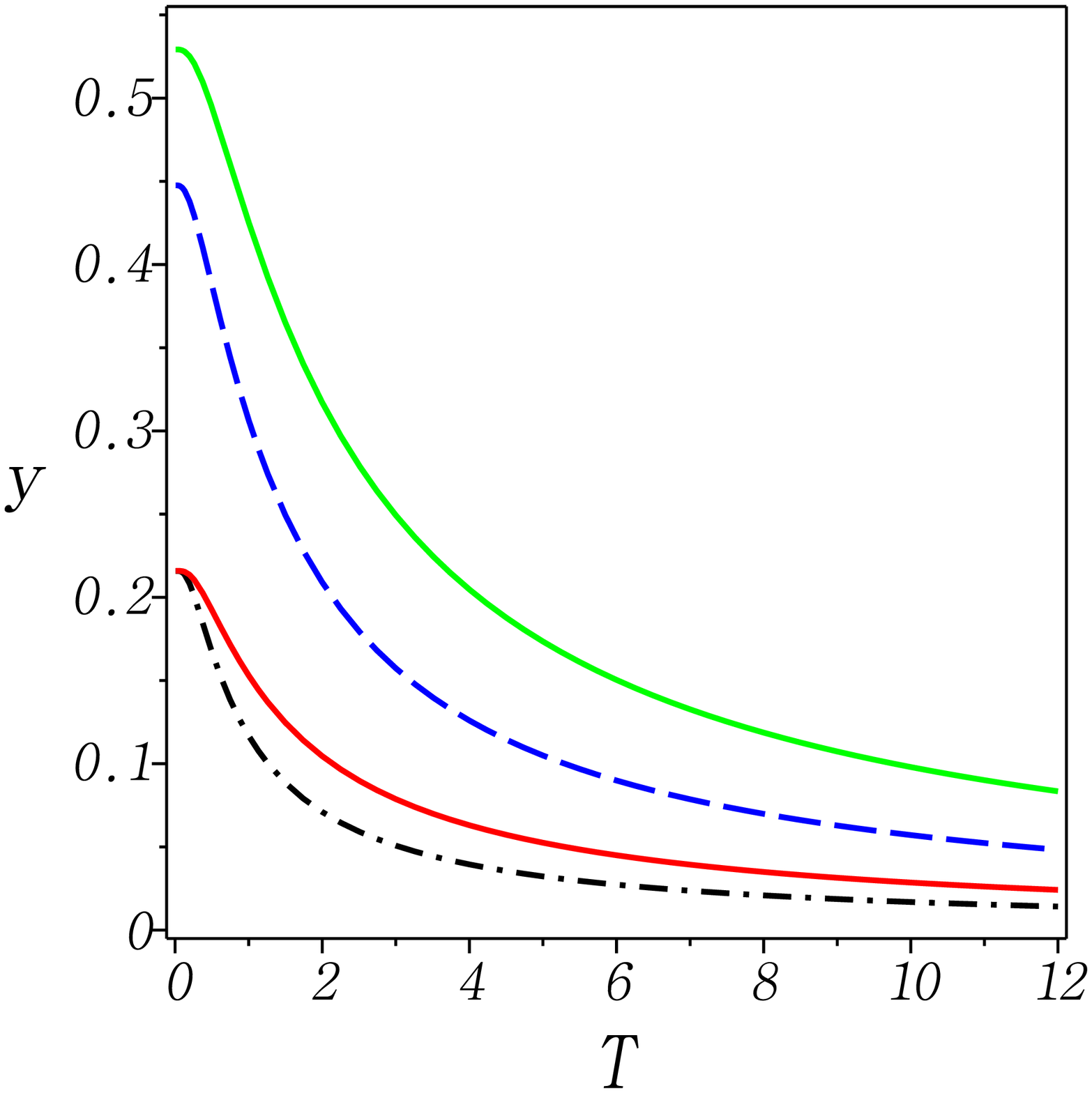}
\caption{\label{fig:fig3}}}
\end{figure}
Fig.~\ref{fig:fig3}: (Color online) The dimensionless energy $y =
\tilde{\mathcal U}_s^*(0)/E_g$ given in
(\ref{eq:energy-drude5-03-1}) with $E_g = \hbar y_0/2$ versus
dimensionless temperature $k_{\mbox{\tiny B}} T/\hbar \bar{y}_0$.
Here we set $M = \hbar = k_{\mbox{\tiny B}} = \bar{y}_0 = 1$. From
top to bottom, 1st) green solid: $\Omega_0 = 3$ and $\bar{\gamma}_0
= 3$ (overdamped); 2nd) blue dash: $\Omega_0 = 1$ and
$\bar{\gamma}_0 = 3$ (overdamped); 3rd) red solid: $\Omega_0 = 3$
and $\bar{\gamma}_0 = 1$ (underdamped); 4th) black dashdot:
$\Omega_0 = 1$ and $\bar{\gamma}_0 = 1$ (underdamped); here
``overdamped'' means $\bar{y}_0 < \bar{\gamma}_0/2$ whereas
``underdamped'' $\bar{y}_0 \geq \bar{\gamma}_0/2$, after
(\ref{eq:useful-relations_1}). With $T \to \infty$, $y \to 0$.
\newpage
\begin{figure}[htb]
\centering\hspace*{-2.5cm}{
\includegraphics[scale=0.75]{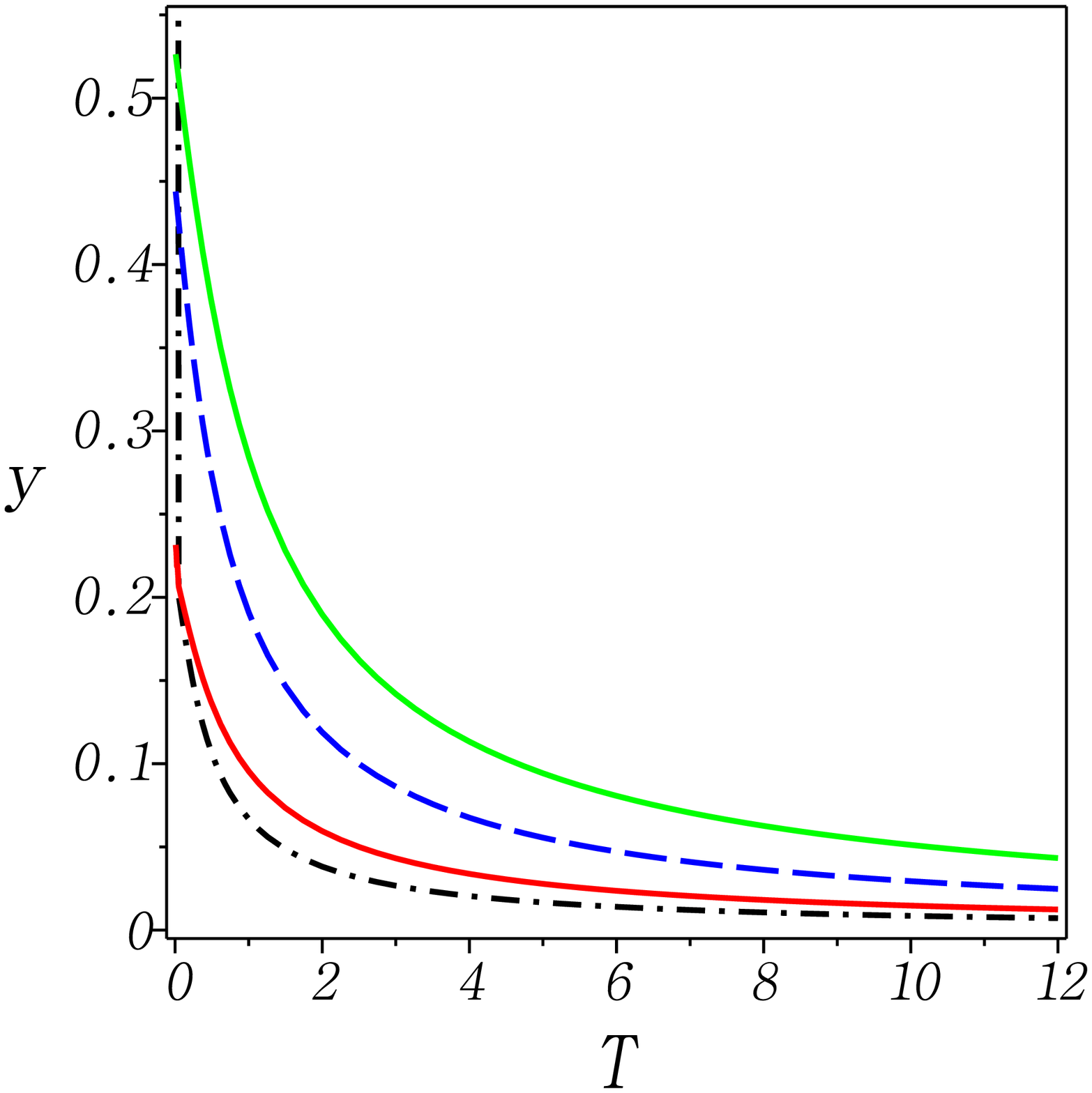}
\caption{\label{fig:fig4}}}
\end{figure}
Fig.~\ref{fig:fig4}: (Color online) The dimensionless free energy $y
= \tilde{\mathcal F}_s^*(0)/E_g$ given in (\ref{eq:final_17-01})
with $E_g = \hbar y_0/2$ versus dimensionless temperature
$k_{\mbox{\tiny B}} T/\hbar \bar{y}_0$. The same as in Fig.
\ref{fig:fig3}.
\newpage
\begin{figure}[htb]
\centering\hspace*{-2.5cm}{
\includegraphics[scale=0.75]{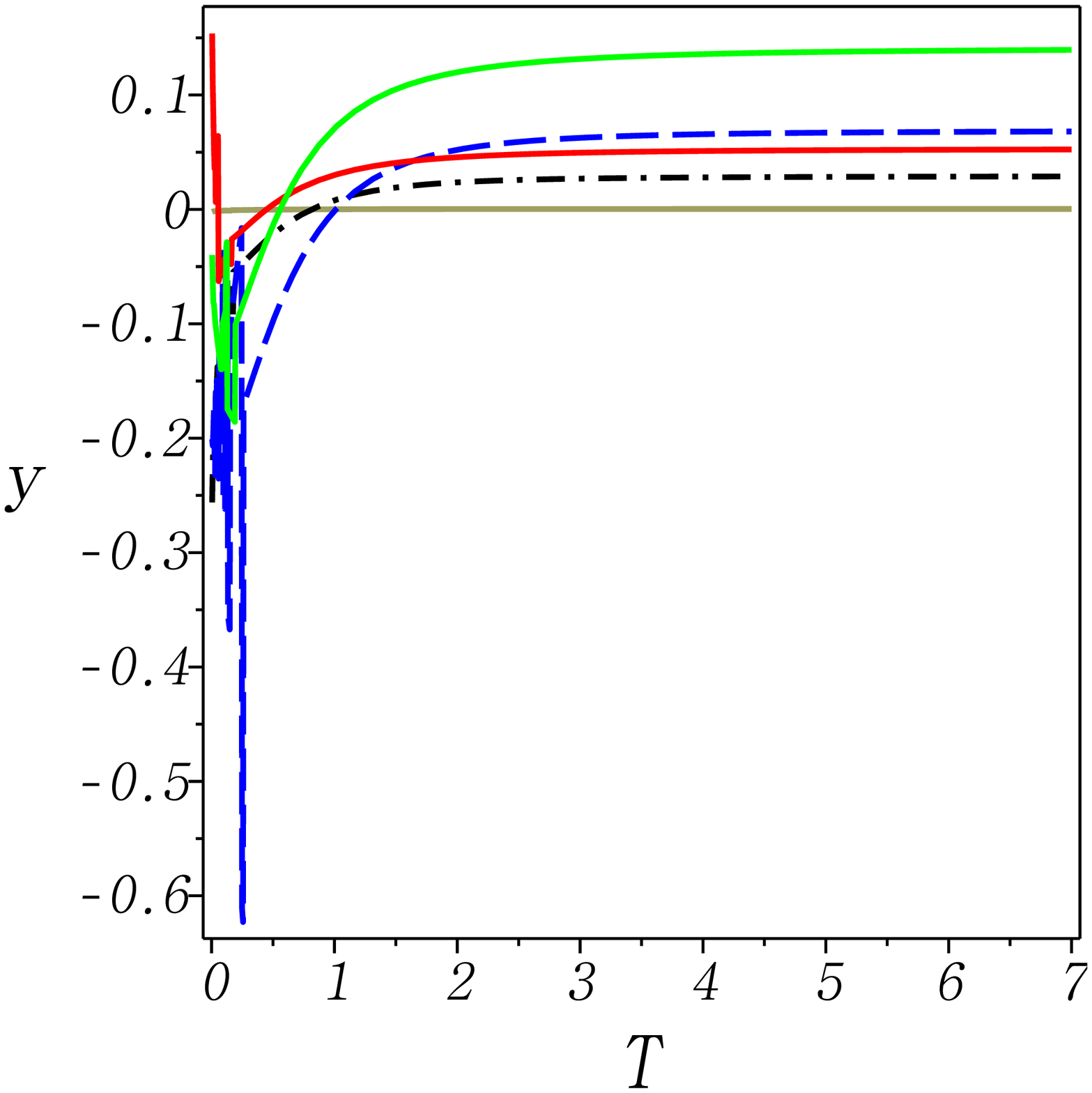}
\caption{\label{fig:fig5}}}
\end{figure}
Fig.~\ref{fig:fig5}: (Color online) The relative error $y =
(\Delta\bar{F}_s(t_{\text{\tiny f}})/\Delta f\{y(t_{\text{\sc
f}}),y_0\}) - 1$ given in (\ref{eq:exact-je-2-1}) versus
dimensionless temperature $k_{\mbox{\tiny B}} T/\hbar y_0$, with
$\Delta f\{y(t_{\text{\sc f}}),y_0\} = f\{\beta,y(t_{\text{\sc
f}})\} - f(\beta,y_0)$ being the leading term on the right-hand side
of (\ref{eq:exact-je-2-1}). Here we set $\hbar = k_{\mbox{\tiny B}}
= y_0 = 1$ and $y(t_{\text{\sc f}}) = 5$. From top to bottom at $T =
7$, 1st) green solid: $\omega_d = 2$ and $\gamma_{\mbox{\tiny o}} =
5$ (strong-coupling limit); 2nd) blue dash: $\omega_d = 7$ and
$\gamma_{\mbox{\tiny o}} = 5$ (Ohmic and strong-coupling limit);
3rd) red solid: $\omega_d = 2$ and $\gamma_{\mbox{\tiny o}} = 1$
(weak-coupling limit); 4th) black dashdot: $\omega_d = 7$ and
$\gamma_{\mbox{\tiny o}} = 1$ (Ohmic and weak-coupling limit); 5th)
khaki solid: $\omega_d = 7$ and $\gamma_{\mbox{\tiny o}} = 0.01$
(Ohmic and vanishingly small coupling limit).
\newpage
\begin{figure}[htb]
\centering\hspace*{-2.5cm}{
\includegraphics[scale=0.75]{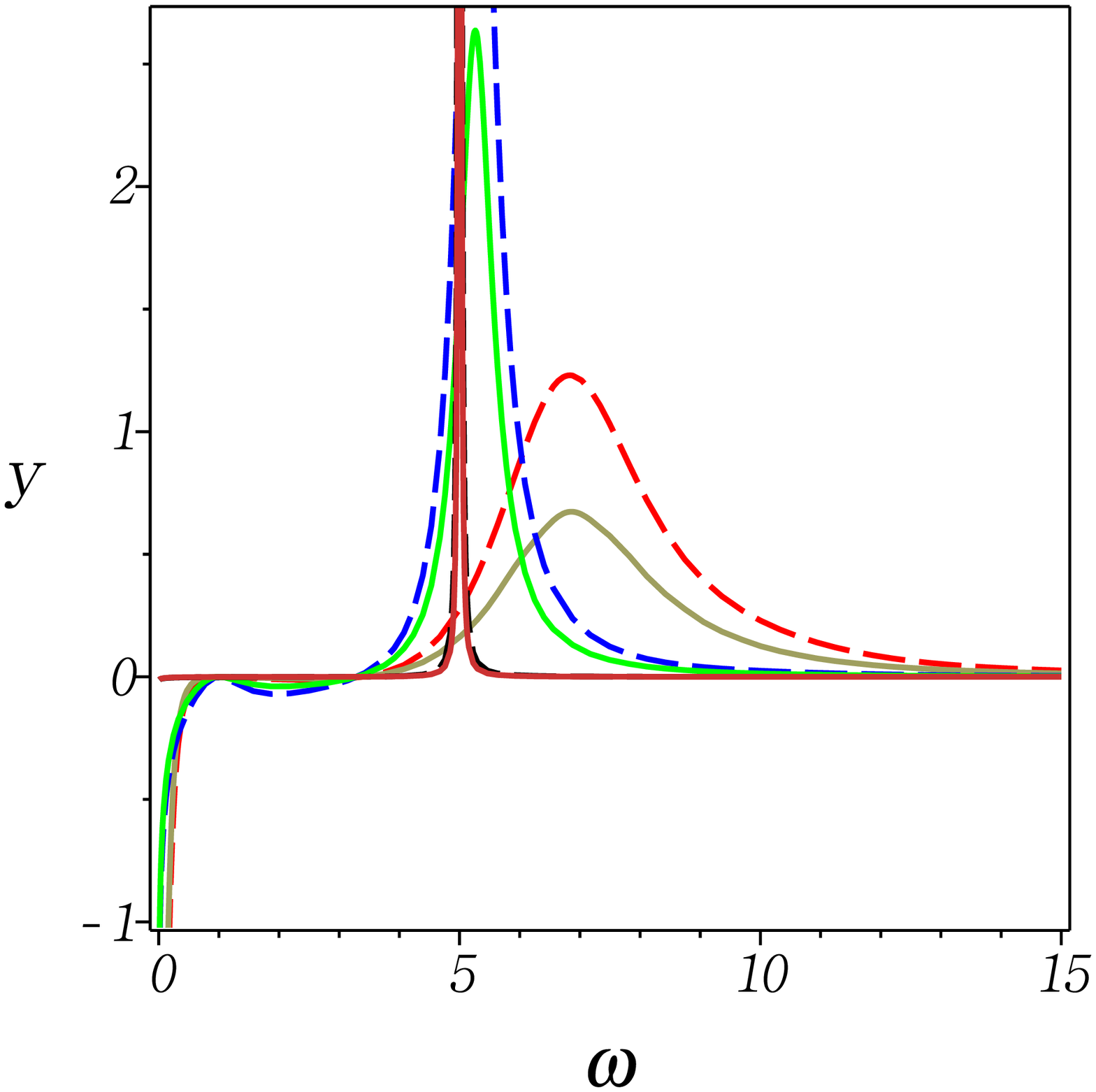}
\caption{\label{fig:fig6}}}
\end{figure}
Fig.~\ref{fig:fig6}: (Color online) The dimensionless quantity $y =
\{\left\<W(\beta,\omega)\right\>_{\scriptscriptstyle{P}} - \Delta
f\}\,\{P_{t_{\text{\tiny f}}}(\omega) - P_0(\omega)\}/(\hbar y_0)$
given in (\ref{eq:exact-je-5}) versus dimensionless frequency
$\omega/y_0$. Here we set $\hbar = k_{\mbox{\tiny B}} = y_0 = 1$ and
$y(t_{\text{\sc f}}) = 5$, and $\omega_d = 7$, as well as $T = 1$
(low-temperature regime). Let $z := \int_0^{\infty} d\omega\cdot
y(\omega)$. (I) Solid lines with the duration $t_{\text{\sc f}} = 5$
(slow change), 1st) orange, with a peak at $\omega = y(t_{\text{\sc
f}})$: $\gamma_{\mbox{\tiny o}} = 0.01$ (vanishingly small coupling)
and $z = 0.3822$; 2nd) green, with maximum value shifted a little to
the right: $\gamma_{\mbox{\tiny o}} = 1$ and $z = 2.3590$; 3rd)
grey, with maximum value shifted further to the right:
$\gamma_{\mbox{\tiny o}} = 5$ (strong coupling) and $z = 1.9298$.
(II) Dash lines with $t_{\text{\sc f}} = 1$ (fast change), in the
same way as in (I), 1st) black: $z = 0.6526$; 2nd) blue: $z =
4.2329$; 3rd) red: $z = 3.6977$. As demonstrated, (1) the smaller
$t_{\text{\sc f}}$, the larger $y$-value, i.e., $1/t_{\text{\sc f}}$
is an irreversibility measure of the process; (2) the $y$-value can
be negative-valued due to its factor $P_{t_{\text{\tiny f}}}(\omega)
- P_0(\omega)$, however, the integral $z$ is non-negative.
\newpage
\begin{figure}[htb]
\centering\hspace*{-2.5cm}{
\includegraphics[scale=0.75]{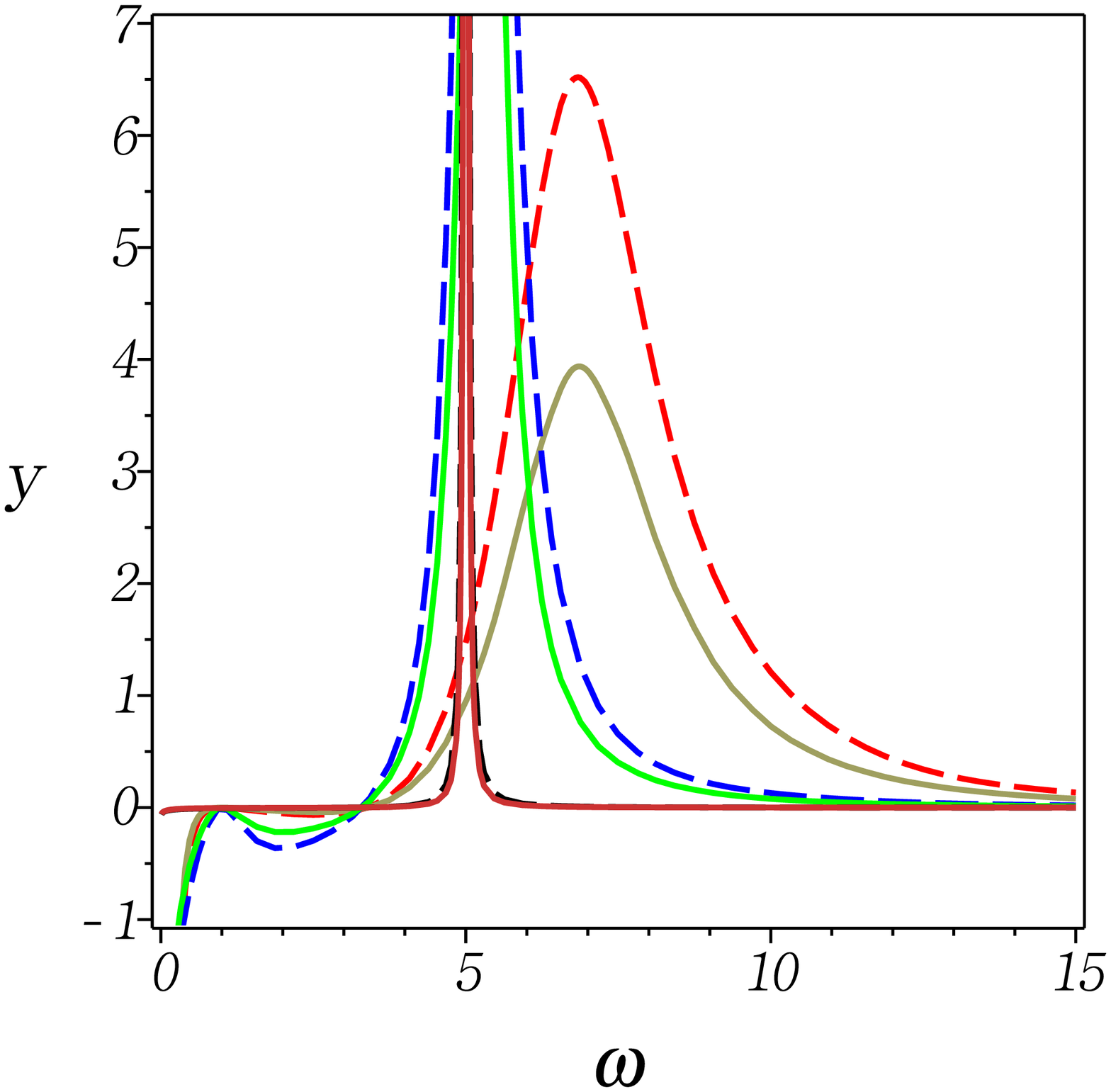}
\caption{\label{fig:fig7}}}
\end{figure}
Fig.~\ref{fig:fig7}: (Color online) The same as in Fig.
\ref{fig:fig6}, but evaluated at $T=5$ (high-temperature regime).
(I) Solid lines with $t_{\text{\sc f}} = 5$ (slow change), 1st)
orange: $z = 2.2343$; 2nd) green: $z = 13.9243$; 3rd) grey: $z =
11.5305$. (II) Dash lines with $t_{\text{\sc f}} = 1$ (fast change),
1st) black: $z = 3.4881$; 2nd) blue: $z = 22.6124$; 3rd) red: $z =
19.7273$.
}

\begin{thebibliography}{1}
\bibitem{ILK13} To verify the normalization conditions for ${\mathcal P}_t^*(\omega)$ and $P_t(\omega)$,
we apply $\int d\omega\cdot a/(\omega^2 + a^2) =
\mbox{arctan}(\omega/a)$ to Eqs.
(\ref{eq:imaginary-susceptibility1}) and
(\ref{eq:distribution-fkt1-11}), respectively.
\end{thebibliography}
\end{document}